\documentclass[aps,prb,twocolumn,showpacs,10pt,groupedaddress]{revtex4-1}
\usepackage{amssymb}
\usepackage{amsmath}
\usepackage[pdftex]{graphicx}
\usepackage{dcolumn}
\usepackage{bm}
\usepackage{textcomp}
\usepackage{color}

\begin{document}

\title{Effective Dirac Hamiltonian for anisotropic honeycomb lattices: optical properties}

\author{M. Oliva-Leyva$1$}
\email{moliva@fisica.unam.mx}
\author{Gerardo G. Naumis$1,2$}
\email{naumis@fisica.unam.mx}

\affiliation{1. Departamento de F\'{i}sica-Qu\'{i}mica, Instituto de
F\'{i}sica, Universidad Nacional Aut\'{o}noma de M\'{e}xico (UNAM),
Apartado Postal 20-364, 01000 M\'{e}xico, Distrito Federal,
M\'{e}xico}
\affiliation{2. School of Physics Astronomy and Computational Sciences, George Mason University, Fairfax, Virginia 22030, USA}


\begin{abstract}

We derive the low-energy Hamiltonian for a honeycomb lattice with anisotropy in the hopping parameters. 
Taking the reported Dirac Hamiltonian for the anisotropic honeycomb lattice, we obtain its optical conductivity tensor and its 
transmittance for normal incidence of linearly polarized light.
Also, we characterize its dichroic character due to the anisotropic optical absorption.
As an application of our general findings, which reproduce the case of uniformly strained
graphene, we study the optical properties of graphene under a nonmechanical distortion. 

\end{abstract}


\maketitle

\section{Introduction}

Among the most unusual properties of graphene, one can cite the linear dispersion relation for electrons and holes at the so-called Dirac 
points.\cite{Neto09} Therefore, at low energies, electrons and holes behave as two-dimensional massless Dirac fermions, which also present 
chiral symmetry. This special property provides for the possibility of observing phenomena such as Klein tunneling,\cite{Katsnelson06,Kim09} 
originally predicted for relativistic particle physics.\cite{Calogeracos} From a practical view point, thinking in the use of graphene for the electronic, 
the unity probability of tunneling through such a barrier, at least for the normal incidence, has resulted a challenge. 

Given the unique mechanical properties of graphene, in particular its striking interval of elastic response,\cite{Lee08,Castellanos} the strain engineering has been 
an alternative to explore the strain-induced modifications of the electronic properties of graphene.\cite{Pereira09b,Guinea12,Wang15,Amorim} Although a theoretical prediction 
of strain-induced opening of a band gap,\cite{Pereira09a,YangLi10,Colombo,Gui15} the most interesting strain-induced effect is the experimental observation of Landau levels signatures to zero 
magnetic field.\cite{Levy,Jiong} As earlier predicted for carbon nanotubes,\cite{Ando02} and subsequently extended to graphene,\cite{Morpurgo,Morozov,Guinea10a} the lattice deformation 
fields can be interpreted in the form of pseudomagnetic fields. Manifestations of such strain-induced pseudomagnetic fields are  
continuously examined, \cite{Salvador13,Falko13,LinHe13,Sandler14,Zenan14,Burgos15,Sandler15,Peeters15,Croy15} even in other materials such as 
transition metal dichalcogenides\cite{Rostami} or Weyl semimetals\cite{Cortijo15}.

In the optical context, strain-induced effects in graphene are significant and open an avenue to potential applications.\cite{Bae13,Shimano,Bin14}
The optical properties of graphene are ultimately determined by its electronic structure which is modified by strain. Needless to say, 
strain produces anisotropy in the electronic dynamics,\cite{Our13} which is traduced in an anisotropic optical conductivity and finally in 
a modulation of the transmittance as a function of the polarization direction.\cite{Pereira10} Such modulation of the transmittance has been observed in
graphene samples under uniaxial strain.\cite{Pereira14} Recently, a theoretical characterization of the transmittance and dichroism was given for graphene 
under an arbitrary uniform strain, e.g., uniaxial, biaxial, and so forth.\cite{Our15a}  

Nowadays, synthetic systems with honeycomb lattices are artificially created to mimic the behavior of Dirac quasiparticles.\cite{Tarruell,Manoharan,Polini} 
The main advantage of these artificial systems is that one can tune in a controlled and independent manner the
hopping of particles between different lattice sites. As a consequence, in such artificial graphene one can observe effects which are induced
by the anisotropy of the hopping parameters, that are not observable in normal graphene under strain. For example, in normal graphene under an 
uniaxial strain, it has been predicted that the Dirac cones can merge.\cite{Pereira09a,Montambaux09a} However, such theoretical prediction requires unrealistic large 
values of strain. However, in artificial graphene of different nature, e.g. of cold atoms or of photonic crystals, the merging of Dirac point 
has been experimentally observed.\cite{Montambaux13} More recently, in electronic artificial graphene, created in a two-dimensional electron gas in a semiconductor 
heterostructure, the merging of Dirac point has been demonstrated for realistic experimental conditions.\cite{Feilhauer}

Undoubtedly, artificial graphene paves new opportunities for studying the physics of Dirac quasiparticles in condensed-matter. Now, 
given the excellent possibility to tune the lattice parameters, it seems necessary to have on hand an effective Dirac
Hamiltonian, which appropriately describes the dynamics of the quasiparticles in anisotropic configurations of the hopping parameters.
In the case of normal graphene under a uniform strain, an effective Dirac Hamiltonian was reported as a function of the strain tensor.\cite{Our13}
However, to the best of our knowledge, the effective Dirac Hamiltonian of artificial graphene, with weak anisotropy of the hopping parameters,
has not been reported in the literature. The main objective of this paper is to give such low-energy Hamiltonian for an anisotropic 
honeycomb lattice (artificial graphene).

\begin{figure*}[t]
{\includegraphics[width=160mm]{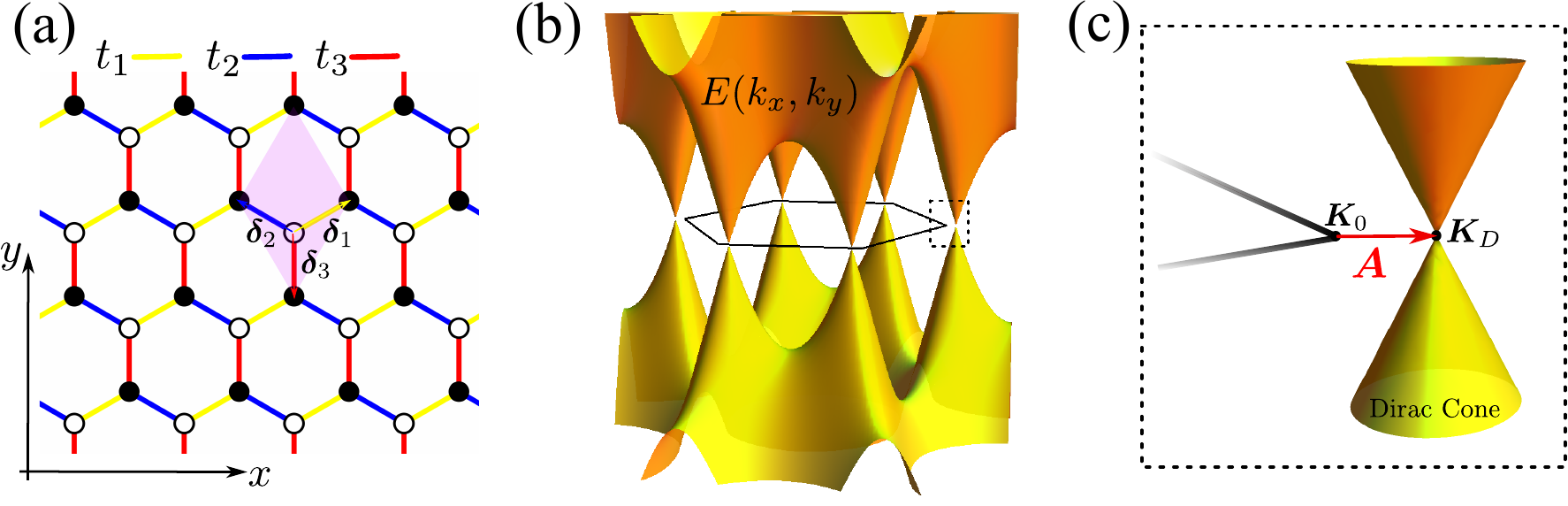}} \caption{\label{fig1} (Color
online) (a) Honeycomb lattice and hoppings. The primitive cell, in light purple, contains a pair of Carbon atoms. 
(b) Dispersion relation $E(\bm{k})$ of the anisotropic honeycomb lattice and its hexagonal first Brillouin zone.
(c) A zoom of the Dirac cone region shows that the Dirac point $\bm{K}_{D}$ is not located at the corner $\bm{K}_{0}$ of the
hexagonal Brillouin zone. The anisotropy-induced Dirac point shift is given by the vector $\bm{A}$.}
\end{figure*}

This paper is organized as follows. In Sec. II, we derive the effective Dirac Hamiltonian 
for a honeycomb lattice with weak anisotropy in the hopping parameters. For this purpose, we start from nearest-neighbor tight-binding model 
 and carry out an expansion around the real Dirac point. The obtained low-energy Hamiltonian is used in Sec. III
to discuss the optical properties of anisotropic honeycomb lattices, which are assumed by having linear response to an external 
oscillating field. In Sec. IV, our findings are particularized to the case of
graphene under a nonmechanical deformation, which can not represented by means of the strain tensor. 
Finally, in Sec. V, our conclusions are given.

\section{Generalized honeycomb lattice}

We are interested in the low-energy Hamiltonian, i.e. the effective Dirac Hamiltonian, of a honeycomb
lattice with anisotropy in the nearest-neighbor hopping parameters. As unstrained graphene, our lattice consists of a triangular Bravais lattice with a 
pair of Carbon atoms (open and filled circles in Fig. 1~(a)) located in its primitive cell. However, we consider that the hoppings between nearest sites 
are dependent on the direction and, in general, are characterized by three hopping parameters $t_{1}$, $t_{2}$ and $t_{3}$ (see Fig. 1~(a)).
Within this nearest-neighbor tight-binding model, one can demonstrate that the Hamiltonian in momentum space can be represented by a 
($2\times2$) matrix of the form\cite{Katsnelson}
\begin{equation}\label{H}
H=-\sum_{n=1}^{3}t_{n}
\begin{pmatrix}
0 & e^{-i\bm{k}\cdot\bm{\delta}_{n}}\\
e^{i\bm{k}\cdot\bm{\delta}_{n}} & 0
\end{pmatrix},
\end{equation}
where $\bm{\delta}_{1},\bm{\delta}_{2},\bm{\delta}_{3}$ are the nearest-neighbor vectors.
Hereafter, we define the the nearest-neighbor vectors as
\begin{equation}
\bm{\delta}_{1}=\frac{a}{2}(\sqrt{3},1), \ \
\bm{\delta}_{2}=\frac{a}{2}(-\sqrt{3},1),\ \
\bm{\delta}_{3}=a(0,-1),
\end{equation}
i.e., we choose the coordinate system $xy$ in a way that the $x$ axis is
along the zigzag direction of the honeycomb lattice (see Fig. 1~(a)). We denote the system $xy$ as the crystalline coordinate system.

From Eq.~(\ref{H}) follows that the dispersion relation is given by two bands:
\begin{equation}\label{DR}
E(\bm{k})=\pm\vert t_{1}e^{i\bm{k}\cdot\bm{\delta}_{1}} + t_{2}e^{i\bm{k}\cdot\bm{\delta}_{2}} + t_{3}e^{i\bm{k}\cdot\bm{\delta}_{3}}\vert,
\end{equation}

As is well documented, for the isotropic case $t_{1,2,3}=t_{0}$, the Dirac points $\bm{K}_{D}$, which are determined by condition $E(\bm{K}_{D})=0$, 
coincide with the corners of the first Brillouin zone. Then, to obtain the Dirac Hamiltonian in this case, one 
can simply expand the Hamiltonian (\ref{H}) around a corner, e.g., $\bm{K}_{0}=(\frac{4\pi}{3\sqrt{3}a},0)$. However, for the considered anisotropic case, the Dirac points do not coincide
with the corners of the first Brillouin zone (as illustrated in Fig. 1~(b,c)).\cite{Pereira09a,YangLi10} In consequence, to derive the Dirac Hamiltonian, one can no longer
expand the Hamiltonian (\ref{H}) around $\bm{K}_{0}$. As recently demonstrated, such expansion around $\bm{K}_{0}$ yields an inappropriate 
Hamiltonian.\cite{Our15b} The appropriate procedure is to \emph{find the position of the Dirac points and carry out the expansion around them}.\cite{Our15b,Volovik14a,Volovik15}
Note that when the hopping anisotropy increases, a gap can appear while the Dirac points disappear. Such effects are characterized by the Hasegawa 
triangular inequalities.\cite{Hasegawa}


\subsection{Effective Dirac Hamiltonian}

Now let us study the effect of a weak anisotropy given by a small perturbation of the hopping parameters. The perturbed hoppings are given by 
\begin{equation}\label{PT}
t_{n}=t_{0}(1+\Delta_{n}),
\end{equation}
on the low-energy description. As expressed above, to derive the proper effective Dirac Hamiltonian it is essential to find 
the position of the Dirac points.\cite{Our15b}
In  Appendix~\ref{AA} we show that from the condition $E(\bm{K}_{D})=0$, up to first order in the parameters $\{\Delta_{n}\}$, one can obtain that $\bm{K}_{D}$ is given by
\begin{equation}\label{KD}
\bm{K}_{D}\approx\bm{K}_{0}+\bm{A},
\end{equation}
where 
\begin{equation}\label{VP}
A_{x}=\frac{1}{3a}(2\Delta_{3}-\Delta_{1}-\Delta_{2}), \ \
\ \ \ A_{y}=\frac{1}{\sqrt{3}a}(\Delta_{1}-\Delta_{2}),
\end{equation}
and $\bm{K}_{0}$ presents valley index $\xi=+1$. As is well known,\cite{Katsnelson,Vozmediano} the shift $\bm{A}$ of the Dirac point
plays the role of an emergent gauge field, similar to a vector potential, when the hopping parameters 
are position-dependent throughout the sample. Thus gauge fields couple with opposite signs to valleys with different index
.\cite{Katsnelson,Vozmediano}  

Once the position of $\bm{K}_{D}$ is found, we expand the Hamiltonian (\ref{H}) around
the Dirac point by means of $\bm{k}=\bm{K}_{D}+\bm{q}$. Following this approach up to first order in the parameters $\{\Delta_{n}\}$,
which is the leading order used throughout the rest of the paper,
we derive that the effective Dirac Hamiltonian results in (see Appendix~\ref{AB}), 
\begin{equation}\label{DH}
H=\hbar
v_{F}\bm{\sigma}\cdot(\bar{\bm{I}}+\bar{\bm{\Delta}})\cdot\bm{q},
\end{equation} 
where $v_{F}=3t_{0}a/2\hbar$ is the Fermi velocity for the unperturbed
honeycomb lattice, $\bm{\sigma}=(\sigma_{x},\sigma_{y})$ are the non-diagonal
Pauli matrices, $\bar{\bm{I}}$ is the $2\times2$ identity matrix and $\bar{\bm{\Delta}}$ is the symmetric matrix
\begin{equation}\label{Delta}
\bar{\bm{\Delta}}=
\begin{pmatrix}
\frac{1}{3}(2\Delta_{1}+2\Delta_{2}-\Delta_{3}) & \frac{1}{\sqrt{3}}(\Delta_{1}-\Delta_{2})\\
\frac{1}{\sqrt{3}}(\Delta_{1}-\Delta_{2}) & \Delta_{3}
\end{pmatrix}.
\end{equation}

Let us note some important remarks about our generalized Hamiltonian (\ref{DH}). First of all, when the three 
hopping parameters $\{t_{n}\}$ are equal to $t_{0}(1+\Delta)$, we have that $\bar{\bm{\Delta}}=\Delta\bar{\bm{I}}$. Then
Eq. (\ref{DH}) reproduces the expected result $\hbar v_{F}(1+\Delta)\bm{\sigma}\cdot\bm{q}$, which is just a renormalization of the Fermi velocity. 
In general, from  Eq. (\ref{DH}) one can recognize a generalized Fermi velocity tensor as
\begin{equation}\label{GV}
\bar{\bm{v}}=v_{F}(\bar{\bm{I}}+\bar{\bm{\Delta}}), 
\end{equation}
whose matrix character is due to the shape of the isoenergetic contours around $\bm{K}_{D}$, which are rotated ellipses. 
Only for the case that $t_{1}=t_{2}$ ($\Delta_{1}=\Delta_{2}$), the Fermi velocity tensor is diagonal with respect to the crystalline coordinate system $xy$
chosen. In this case, the principal axes of the isoenergetic ellipses are collinear with the $xy$ axes. 
Moreover, the general character of Eq. (\ref{DH}) enables to estimate the variation effects of the hopping parameters for graphene
under a spatially uniform strain. For the last case, to first order in the strain tensor $\tilde{\bm{\epsilon}}$, the hopping parameters are 
approximated by $t_{n}\approx t_{0}(1-\beta\bm{\delta}_{n}\cdot\bar{\bm{\epsilon}}\cdot\bm{\delta}_{n}/ a^{2})$, where $\beta$ is the electron Gr\"{u}neisen
parameter. Thus, $\Delta_{n}=-\beta\bm{\delta}_{n}\cdot\bar{\bm{\epsilon}}\cdot\bm{\delta}_{n}/ a^{2}$ and as a consequence, one obtain that 
for uniformly strained graphene, $\bar{\bm{\Delta}}=-\beta\bar{\bm{\epsilon}}$. However, to write the effective Dirac Hamiltonian for uniformly 
strained graphene, it should also take into account the deformation of the  lattice vectors.\cite{Our13} 

It is important to emphasize that the expression (\ref{Delta}) for the matrix $\bar{\bm{\Delta}}$ is referred to the crystalline 
coordinate system $xy$. In general, if one choose an arbitrary coordinate system $x'y'$, rotated at an angle $\phi$ respect to the system $xy$,
then the new components of $\bar{\bm{\Delta}}$ can be found by means of the transformation rules of a second order Cartesian tensor. 
In other words, $\bar{\bm{\Delta}}$ is a second order Cartesian tensor, whose explicit form (\ref{Delta}) is given with respect to the crystalline 
coordinate system $xy$.

\section{Optical conductivity}

Let us now study the optical properties of those
anisotropic electronic honeycomb fermionic lattices that exhibit linear response to an external electric
field of frequency $w$. For this purpose, we firstly obtain the optical conductivity tensor
$\bar{\sigma}_{ij}(w)$ by combining the Hamiltonian (\ref{DH}) and the
Kubo formula. Following the representation used in
Refs.~[\cite{Ziegler06,Ziegler07}], the optical conductivity $\sigma_{ij}(w)$ can be written as a double integral with
respect to two energies $E$, $E'$:
\begin{eqnarray} \label{C}
\bar{\sigma}_{ij}(w)=\frac{i}{\hbar}\int\int \text{Tr}\{ j_{i}\delta(H-E')j_{j} \delta(H-E)\} \nonumber\\
\qquad\times\frac{1}{E-E'+w-i\alpha}\frac{f(E)-f(E')}{E-E'}\text{d}
E\text{d} E',
\end{eqnarray}
where $f(E)=(1+\exp[E/(k_{B}T)])^{-1}$ is the Fermi function at
temperature $T$ and $j_{l}=-i e [H,r_{l}]$ is the current operator
in the $l$-direction, with $l=x,y$.

To calculate the integral (\ref{C}) it is convenient to carry out the change of variables
\begin{equation}\label{CV}
\bm{q}=(\bar{\bm{I}}+\bar{\bm{\Delta}})^{-1}\cdot\bm{q}^{*}
\end{equation} 

In the new variables
$(q_{x}^{*},q_{y}^{*})$, the Hamiltonian (\ref{DH}) becomes 
$H=\hbar v_{F}\bm{\sigma}\cdot\bm{q}^{*}$, corresponding to an unperturbed honeycomb lattice, as unstrained graphene. 
On the other hand, the current operator components transform as
\begin{eqnarray}\label{jx}
j_{x}&=&-i e[H,r_{x}]=e\frac{\partial H}{\partial q_{x}}, \nonumber\\
&=&e\left(\frac{\partial H}{\partial q_{x}^{*}} \frac{\partial
q_{x}^{*}}{\partial q_{x}} + \frac{\partial H}{\partial q_{y}^{*}}
\frac{\partial q_{y}^{*}}{\partial q_{x}}\right), \nonumber\\
&=&(1 + \bar{\Delta}_{xx})j_{x}^{*}
+ \bar{\Delta}_{xy}j_{y}^{*},
\end{eqnarray}
and analogously
\begin{equation}\label{jy}
j_{y}=(1 + \bar{\Delta}_{yy})j_{y}^{*}
+ \bar{\Delta}_{xy}j_{x}^{*},
\end{equation}
where $j_{x}^{*}=e(\partial H/\partial q_{x}^{*})$ and
$j_{y}^{*}=e(\partial H/\partial q_{y}^{*})$ are the current
operator components for the case of the unperturbed honeycomb lattice. 

Then, substituting Eqs.~(\ref{jx}) and (\ref{jy}) into
Eq.~(\ref{C}), we obtain
\begin{eqnarray}
\bar{\sigma}_{xx}(w)&\simeq&(1+2\bar{\Delta}_{xx})J\sigma_{0}(w),\label{sxx}\\
\bar{\sigma}_{yy}(w)&\simeq&(1+2\bar{\Delta}_{yy})J\sigma_{0}(w),\\
\bar{\sigma}_{xy}(w)&=&\bar{\sigma}_{yx}(w)\simeq 2\bar{\Delta}_{xy}J\sigma_{0}(w),\label{sxy}
\end{eqnarray}
where $J$ is the Jacobian determinant of the transformation (\ref{CV}) and $\sigma_{0}(w)$
is the optical conductivity of the unperturbed honeycomb lattice.  
Note that as an explicit expression for $\sigma_{0}(w)$ one can use the reported optical conductivity of unstrained 
graphene.\cite{Ziegler07,Gusynin07,Stauber08} 

Finally, from Eqs.~(\ref{sxx})--(\ref{sxy}) it follows that the optical conductivity tensor
for the anisotropic honeycomb lattice 
results in,
\begin{equation}\label{CVF}
\bar{\bm{\sigma}}(w)\simeq\sigma_{0}(w)(\bar{\bm{I}} + 2\bar{\bm{\Delta}} - \text{Tr}(\bar{\bm{\Delta}})\bar{\bm{I}}).
\end{equation}

In other words, a Dirac system described by the generalized Hamiltonian (\ref{DH}) presents an anisotropic optical response given 
by Eq.~(\ref{CVF}), independently of the expression of matrix $\bar{\bm{\Delta}}$. Now substituting Eq.~(\ref{Delta}) into 
Eq.~(\ref{CVF}) we obtain the explicit form of the optical conductivity tensor for the anisotropic honeycomb lattice,
respect to the crystalline coordinate system $xy$.


\subsection{Dichroism and Transmittance}

The anisotropy of the optical absorption yields two effects: dichroism and modulation of the transmittance as a function of the 
polarization direction. To examine such effects, let us consider normal incidence of linearly polarized light between two dielectric media 
separated by our anisotropic honeycomb lattice, as illustrated in Fig.~2(a). From the boundary conditions for the electromagnetic field on 
the interface between both media, one can obtain that the electric fields of the incident and transmitted waves, $\bm{E}_{i}$ and $\bm{E}_{t}$
respectively, are related by\cite{Our15a}
\begin{equation}\label{Di}
 \bm{E}_{i}=\frac{1}{2}\sqrt{\frac{\mu_{1}}{\epsilon_{1}}}
 \left(\left(\sqrt{\frac{\epsilon_{1}}{\mu_{1}}}+\sqrt{\frac{\epsilon_{2}}{\mu_{2}}}\right)\bar{\bm{I}}+\bar{\bm{\sigma}}\right)\cdot\bm{E}_{t},
\end{equation}
where $\epsilon_{1,2}$  are the electrical permittivities  and $\mu_{1,2}$, the
magnetic permeabilities. Note that, in general, the anisotropy of the conductivity $\bar{\bm{\sigma}}$ produces that $\bm{E}_{i}$ and 
$\bm{E}_{t}$ are not collinear (see Fig.~2(b)). Analogously, the magnetic fields, $\bm{H}_{i}$ and $\bm{H}_{t}$, fulfill the same relation.

Now from Eq.~(\ref{Di}) the calculation of the transmittance is straightforward:\cite{Our15a}
\begin{eqnarray}\label{T}
 T(\theta_{i})&=&\frac{|\bm{E}_{t}\times\bm{H}_{t}|}{|\bm{E}_{i}\times\bm{H}_{i}|}=
 \frac{\sqrt{\epsilon_{2}/\mu_{2}}|\bm{E}_{t}|^{2}}{\sqrt{\epsilon_{1}/\mu_{1}}|\bm{E}_{i}|^{2}},\nonumber\\
 &\approx& T_{0}\left(1-\frac{2\sqrt{\mu_{1}\mu_{2}}}{\sqrt{\epsilon_{1}\mu_{2}}+
 \sqrt{\epsilon_{2}\mu_{1}}}\bm{e}_{i}^{\intercal}\cdot\mbox{Re}\bar{\bm{\sigma}}\cdot\bm{e}_{i}\right), 
\end{eqnarray}
where $T_{0}$ is the transmittance for normal incidence between two media in absence of the anisotropic honeycomb lattice as interface and 
$\bm{e}_{i}^{\intercal}=(\cos\theta_{i},\sin\theta_{i})$, with $\theta_{i}$ being the incident polarization angle $\theta_{i}$. 

To illustrate even more clearly the dichroism and the modulation of $T(\theta_{i})$ induced by the anisotropic honeycomb lattice,
it is convenient to consider that the chemical potential equals zero for the lattice. In consequence, for the domain of
infrared and visible frequencies, in Eq.~(\ref{CVF}) $\sigma_{0}(w)$ can be replaced by the universal and frequency-independent value 
$e^{2}/(4\hbar)$.\cite{Stauber08,Nair08} Additionally, if both media are vacuum, then from Eqs.(\ref{CVF})-(\ref{T}) we obtain that 
\begin{eqnarray}
 \theta_{t}-\theta_{i}&\approx&\frac{\pi\alpha}{2}A\sin(2\theta_{i}-\theta_{0}),\ \ \ \text{(in radians)},\label{DiS}\\
 T(\theta_{i})&\approx& 1-\pi\alpha\bigl(1 + A\cos(2\theta_{i}-\theta_{0}))\label{TS},
\end{eqnarray}
where
\begin{equation}
 A=\sqrt{(\bar{\Delta}_{xx}-\bar{\Delta}_{yy})^{2} + 4 \bar{\Delta}_{xy}^{2}}=
 \sqrt{(\mbox{Tr}\bar{\bm{\Delta}})^{2}-4\mbox{Det}\bar{\bm{\Delta}}},
\end{equation}
$\sin\theta_{0}=2\bar{\Delta}_{xy}/A$ ($\cos\theta_{0}=(\bar{\Delta}_{xx}-\bar{\Delta}_{yy})/A$) and $\alpha$ is the fine-structure constant. 

\begin{figure}[t]
\includegraphics[width=8.cm]{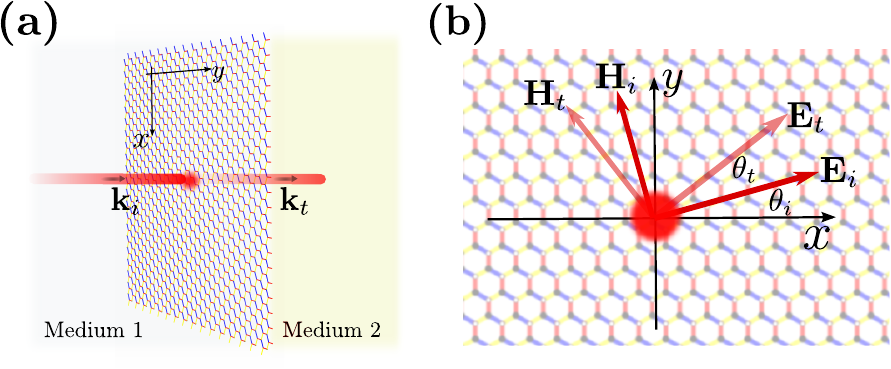}
\caption{\label{fig2} (Color online) (a) Scattering problem for normal incidence between two media with the anisotropic honeycomb lattice
separating them. The $z$ direction is chosen along the propagation of the electromagnetic wave. $\bm{k}_{i}$ and $\bm{k}_{t}$ represent
the wave vectors of the incident and transmitted waves, respectively. (b) Schematic representation of the dichroism 
induced by the anisotropic absorption of the honeycomb lattice. The electromagnetic fields lie in the lattice plane.}
\end{figure}

It is immediate to verify that, for $\bar{\bm{\Delta}}=0$, the dichroism disappears and $T(\theta_{i})$ reduces to $1-\pi\alpha$ $(\approx97.7\%)$, 
which is the transmittance of unstrained graphene.\cite{Nair08} Expressions (\ref{DiS}) and (\ref{TS}) clearly show $\pi$-periodic modulations of the dichroism and 
transmittance respect to the incident polarization angle $\theta_{i}$, which is due to the physical equivalence between $\theta_{i}$ and 
$\theta_{i}+\pi$, for normal incidence of linearly polarized light. 
From Eq.~(\ref{DiS}) follows that the principal directions of $\bar{\bm{\Delta}}$ can be determined by monitoring the polarization angles 
$\theta_{i}$ for which the incident and transmitted polarizations coincide. At the same time, Eq.~(\ref{TS}) shows that the principal 
directions of $\bar{\bm{\Delta}}$ can be determined by measuring the polarization angles $\theta_{i}$ for which the transmittance takes its
minimum or maximum values. Also, it is important to note that while the phase $\theta_{0}$ of both modulations is dependent on the
coordinate system orientation, the amplitude $A$ is independent on any orientation of the coordinate system, as physically expected, 
because $A$ is given as a function of the invariants $\mbox{Tr}\bar{\bm{\Delta}}$  and $\mbox{Det}\bar{\bm{\Delta}}$. 

So far, our discussion on dichroism and transmittance is general for an anisotropic Dirac system described by Eq.~(\ref{DH}).
By substituting Eq.~(\ref{Delta}) into Eqs.~(\ref{DiS},\ref{TS}), one obtains the explicit form of such effects
for the anisotropic honeycomb lattice with respect to the crystalline coordinate system $xy$.


\section{Application}

\begin{figure}[t]
\includegraphics[width=6.cm]{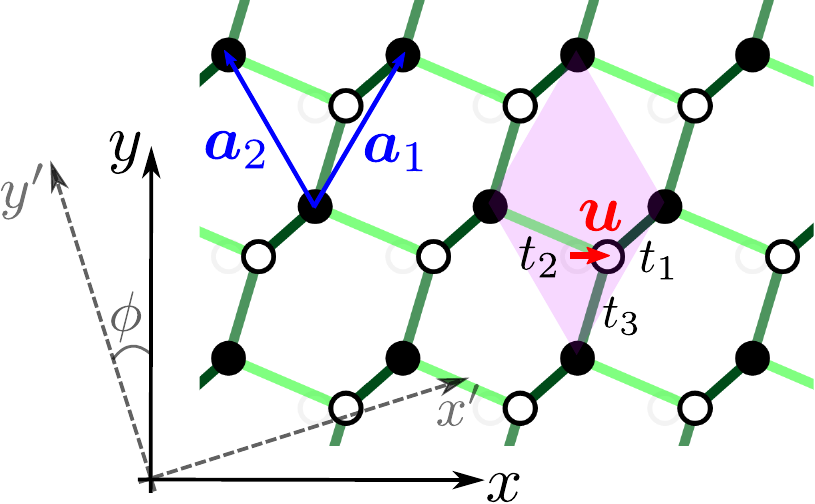}
\caption{\label{fig3}(Color online) Scheme of the modified graphene lattice. The red arrow represents the displacement $\bm{u}$ of the open-circles sublattice.}
\end{figure}

As an application of the previous results, we consider a deformation of graphene lattice, as illustrated in Fig.~\ref{fig3}(a).
Here the primitive vectors $\bm{a}_{1}$ and $\bm{a}_{2}$ remain undeformed, while the atom of basis,
denoted by an open circle, is displaced a vector $\bm{u}$  in an arbitrary direction. This deformation of graphene lattice is basically a displacement,
given by $\bm{u}$, of the open-circles sublattice  respect to the filled-circles sublattice.  
A possible scenario for such deformation could occur in graphene grown on substrate with an appropriate
combination of lattice mismatch between the two crystals.\cite{Pereira14,Seung15}

The new nearest-neighbor vectors $\bm{\delta}_{n}'$ are related to the unstrained nearest-neighbor vectors
$\bm{\delta}_{n}$ by means of $\bm{\delta}_{n}'=\bm{\delta}_{n} - \bm{u}$. However, it is worth mentioning that
the reciprocal lattice of our modified graphene lattice remains undeformed because 
the direct lattice, determined by $\bm{a}_{1}$ and $\bm{a}_{2}$, is undistorted. The last is a notable difference 
with the case of strained graphene by means of mechanical stress, for which the lattice vectors are deformed. 

As in Sec. II, if we begin from a nearest-neighbor approach, it is easy to demonstrate that the dispersion relation for this
modified graphene lattice reads
\begin{eqnarray}\label{DRMG}
E(\bm{k})&=&\pm\vert t_{1}e^{i\bm{k}\cdot(\bm{\delta}_{1}-\bm{u})} + t_{2}e^{i\bm{k}\cdot(\bm{\delta}_{2}-\bm{u})} + 
t_{3}e^{i\bm{k}\cdot(\bm{\delta}_{3}-\bm{u})}\vert,\nonumber\\
&=&\pm\vert t_{1}e^{i\bm{k}\cdot\bm{\delta}_{1}} + t_{2}e^{i\bm{k}\cdot\bm{\delta}_{2}} + t_{3}e^{i\bm{k}\cdot\bm{\delta}_{3}}\vert,
\end{eqnarray}
where we characterize the variation of the hopping parameters in the usual form:
$t_{n}=t_{0}\exp[-\beta(|\bm{\delta}_{n}'|/a - 1)]$.\cite{Pereira09a}  Note that Eq.~(\ref{DRMG}) coincides with Eq.~(\ref{DR}), so the modified graphene 
lattice can be considered as a particular case of the generalized honeycomb lattice examined in Sec. II. Therefore, now we can particularize 
all general previous results for the modified graphene lattice.

\emph{Effective Dirac Hamiltonian}: Writing the variation of the hopping parameters to first order in $\bm{u}$, one get
\begin{eqnarray}
 t_{n}&=&t_{0}\exp[-\beta(|\bm{\delta}_{n}-\bm{u}|/a - 1)],\nonumber\\
 &\approx&t_{0}(1+\beta\bm{\delta}_{n}\cdot\bm{u}/a^{2}),
\end{eqnarray}
thus, for this case one can identify from Eq.(\ref{PT}) that $\Delta_{n}=\beta\bm{\delta}_{n}\cdot\bm{u}/a^{2}$. Consequently, 
from Eqs.(\ref{DH}) and (\ref{Delta}), the effective Dirac Hamiltonian of the modified graphene lattice results in,
\begin{equation}\label{DHMG}
H=\hbar
v_{F}\bm{\sigma}\cdot(\bar{\bm{I}}+\bar{\bm{\Delta}}^{u})\cdot\bm{q},
\end{equation}
where the matrix $\bar{\bm{\Delta}}_{u}$ dependents on the components of the vector $\bm{u}$ as
\begin{equation}\label{DeltaU}
\bar{\bm{\Delta}}^{u}=\frac{\beta}{a}
\begin{pmatrix}
u_{y} & u_{x}\\
u_{x} & -u_{y}
\end{pmatrix}.
\end{equation}

It is immediate to verify that for $\bm{u}=0$ one recover the case of unstrained graphene. Note that $\mbox{Tr}\bar{\bm{\Delta}}^{u}=0$, which is 
because the studied deformation does not vary the area of graphene sample. This fact is analogous to having a pure shear strain.  

The form (\ref{DeltaU}) of the tensor $\bar{\bm{\Delta}}^{u}$ is referred to the crystalline coordinate system $xy$. Now let us give its general
expression respect to an arbitrary coordinate system $x'y'$, which is inclined to $xy$ at an angle $\phi$. Using
the transformation rules of a second order Cartesian tensor we find 
\begin{eqnarray}\label{GDeltaU}
 \bar{\Delta}^{u}_{x'x'}&=&-\bar{\Delta}^{u}_{y'y'}=\frac{\beta}{a}(u_{y'}\cos3\phi + u_{x'}\sin3\phi),\\
 \bar{\Delta}^{u}_{x'y'}&=&\bar{\Delta}^{u}_{y'x'}=\frac{\beta}{a}(-u_{y'}\sin3\phi + u_{x'}\cos3\phi),
\end{eqnarray}
where $u_{x'}$ and $u_{y'}$ are the components of the vector $\bm{u}$ respect to the system $x'y'$. 
These expressions for the tensor $\bar{\bm{\Delta}}^{u}(\phi)$ exhibit a clear periodicity of $2\pi/3$ in $\phi$,
which reflects the trigonal symmetry of the underlying honeycomb lattice.

\begin{figure}[t]
\includegraphics[width=8.cm]{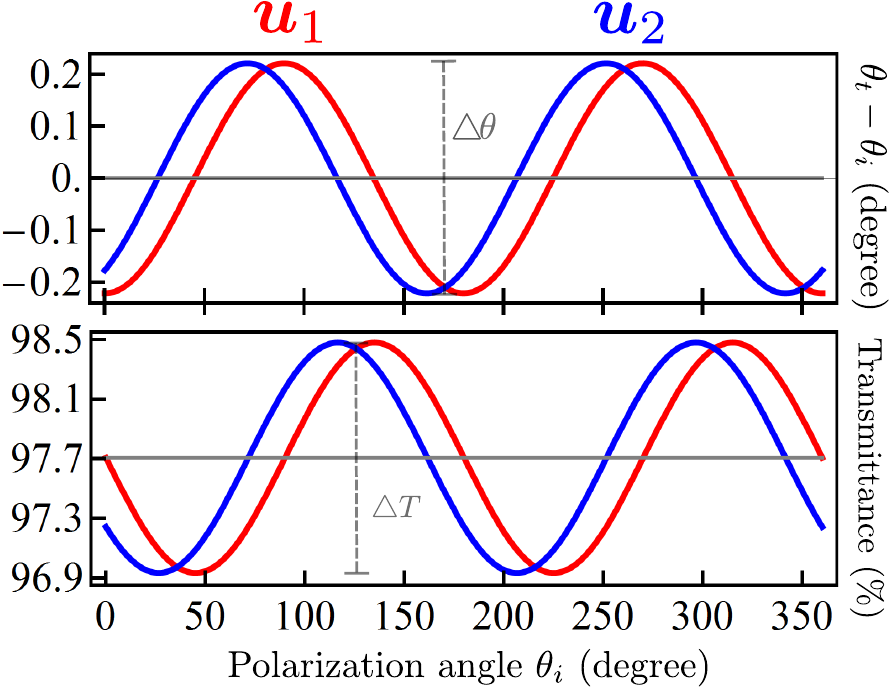}
\caption{\label{fig4}(Color online)
Rotation of the transmitted field (top panel) and transmittance (bottom panel) as a function of the incident polarization angle for two
different deformations. The red curves correspond to the displacement $\bm{u}_{1}/a=(0.05,0)$, while the blue curves correspond to 
the displacement $\bm{u}_{2}/a=(0.04,0.03)$.}
\end{figure}


\emph{Optical properties}: From Eqs. (\ref{CVF}) and (\ref{DeltaU}), the optical conductivity for the modified graphene lattice
immediately follows as
\begin{eqnarray}
 \bar{\bm{\sigma}}^{u}(w)&=&\sigma_{0}(w)(\bar{\bm{I}} + 2\bar{\bm{\Delta}}^{u}),\nonumber\\
 &=&\sigma_{0}(w)
\begin{pmatrix}
1+2\beta u_{y}/a & 2\beta u_{x}/a\\
2\beta u_{x}/a & 1 - 2\beta u_{y}/a
\end{pmatrix},
\end{eqnarray}
with respect to the crystalline coordinate system $xy$. At the same time, from Eqs. (\ref{DiS}) and (\ref{TS}) one obtains that, for normal incidence of
linearly polarized light, the dichroism and the transmittance are characterized by
\begin{eqnarray}
 \theta_{t}-\theta_{i}&\approx&\frac{\pi\alpha\beta}{a}(u_{y}\sin2\theta_{i} - u_{x}\cos2\theta_{i}),\label{DiSU}\\
 T(\theta_{i})&\approx& 1-\pi\alpha -\frac{2\pi\alpha\beta}{a}(u_{y}\cos2\theta_{i}+u_{x}\sin2\theta_{i}), \label{TSU}
\end{eqnarray}
where the incident polarization angle $\theta_{i}$ is measured respect to the $x$ axis of the crystalline coordinate system.  

In Fig.~\ref{fig4}, we display the evaluated expressions (\ref{DiSU}) and (\ref{TSU}) for two different displacements, 
$\bm{u}_{1}/a=(0.05,0)$ and $\bm{u}_{2}/a=(0.04,0.03)$. Such deformations present the same modulation amplitude
either for the rotation of the transmitted field (dichroism) as for the transmittance. The reason is simple. 
From  (\ref{TSU}) it follows that the transmittance modulation amplitude   $\triangle T$ is determined by the module
of the vector $\bm{u}$: $\triangle T=2\pi\alpha\beta|\bm{u}|/a$. Note that $|\bm{u}_{1}|=|\bm{u}_{2}|$, therefore,
$\triangle T_{1}$= $\triangle T_{2}$. The analogous argument is valid for the modulation amplitude of the rotation of the transmitted field:
$\triangle \phi=\pi\alpha\beta|\bm{u}|/a$.

Nowadays, measurements of transmittance can be used to characterize the deformation state in graphene sample.\cite{Pereira14,Our15a}
 
To end, let us point out how Eq. (\ref{TSU}) serves to characterize the strain state of our modified graphene lattice by means of two measurements of transmittance. 
Measuring the transmittance at $\theta_{i}=0$ and $\theta_{i}=\pi/4$, from Eq. (\ref{TSU}) one obtain that
\begin{eqnarray}
u_{x}&=&\frac{a}{2\pi\alpha\beta}(1-\pi\alpha -T(\pi/4)),\\
u_{y}&=&\frac{a}{2\pi\alpha\beta}(1-\pi\alpha -T(0)),
\end{eqnarray}
and of this manner, the strain state can be determined.

\section{Conclusion}

In summary, starting from a nearest-neighbor tight-binding model we derived the effective Dirac Hamiltonian of an anisotropic honeycomb lattice, 
beyond strained normal graphene.
This general Hamiltonian results in a useful tool for studying the anisotropic dynamics of Dirac quasiparticles in artificial graphene. Moreover, it is an 
excellent starting point for obtaining the effective Dirac Hamiltonian in the case of position-dependent anisotropy of the 
honeycomb lattice.\cite{Our15b} We also obtained the optical conductivity tensor of the anisotropic honeycomb lattice and showed how
such anisotropic optical absorption produces a modulation of the transmittance and of the dichroism as a function of the incident polarization angle. 
Our findings could provide a platform to characterize the anisotropy in electric artificial graphene by means of optical measurements.
At the same time, they could be used for tailoring the optical properties of electric artificial graphene.

\begin{acknowledgments}
This work was supported by UNAM-DGAPA-PAPIIT, project IN-$102513$. M.O.L
acknowledges support from CONACYT (Mexico). G.G. Naumis thanks a PASPA scholarship for a
sabbatical leave at the George Mason University.
\end{acknowledgments}

\appendix

\section{}\label{AA}

Here we provide the derivation of the expressions~(\ref{VP}) for the shift vector $\bm{A}$ of the Dirac point $\bm{K}_{D}$.

The condition $E(\bm{K}_{D})=0$, which defines the Dirac points $\bm{K}_{D}$, can be equivalently rewritten as 
\begin{equation}\label{Ec1}
 \sum_{n=1}^{3}t_{n}e^{i\bm{K}_{D}\cdot\bm{\delta}_{n}}=0.
\end{equation}

In the isotropic case, $t_{n}=t_{0}$, the Dirac points coincide with the corners of the first Brillouin zone, in particular,
$\bm{K}_{D}=\bm{K}_{0}=(\frac{4\pi}{3\sqrt{3}a},0)$, with $\bm{K}_{0}$ being a corner with valley index $+1$. For the anisotropic case, 
$t_{n}=t_{0}(1 + \Delta_{n})$, the Dirac points do not coincide with the corners of the first Brillouin zone, in particular,
$\bm{K}_{D}\neq\bm{K}_{0}$. Then, one can propose the position of $\bm{K}_{D}$ in the form
\begin{equation}\label{Ec2}
 \bm{K}_{D}=\bm{K}_{0} + \bm{A} + \mathcal{O}(\Delta_{n}^{2}),
\end{equation}
where the unknown shift $\bm{A}$ will be looked for as a lineal combination on the parameters $\{\Delta_{n}\}$. Now,
substituting Eq.~(\ref{Ec2}) into Eq.~(\ref{Ec1}) results 
\begin{eqnarray}
 \sum_{n=1}^{3}t_{0}(1 + \Delta_{n})e^{i(\bm{K}_{0} + \bm{A} + \mathcal{O}(\Delta_{n}^{2}))\cdot\bm{\delta}_{n}}&=&0,\nonumber \\
 \sum_{n=1}^{3}(1 + \Delta_{n}+ i\bm{A}\cdot\bm{\delta}_{n} + \mathcal{O}(\Delta_{n}^{2}))e^{i\bm{K}_{0}\cdot\bm{\delta}_{n}}&=&0,\nonumber \\
 -3aA_{x}-\Delta_{1}-\Delta_{2}+ 2\Delta{3}\ \ \ \ \ \ \ \ \ \ \ \ \ \ \ \ \ &&\nonumber\\ 
 + i (-3 a A_{y} + \sqrt{3}\Delta_{1} - \sqrt{3}\Delta_{2}) + \mathcal{O}(\Delta_{n}^{2})&=&0.\label{Ec3}
\end{eqnarray}

Thus, one obtain that the shift $\bm{A}$ is given by 
\begin{equation}
A_{x}=\frac{1}{3a}(2\Delta_{3}-\Delta_{1}-\Delta_{2}), \ \
\ \ \ A_{y}=\frac{1}{\sqrt{3}a}(\Delta_{1}-\Delta_{2}),
\end{equation}
which is our Eq.~(\ref{VP}). Following a similar calculation, 
the Dirac point shift respect to the corner $\bm{K}_{0}^{'}=(-\frac{4\pi}{3\sqrt{3}a},0)$, with valley index $-1$,
results in $-\bm{A}$.

\section{}\label{AB}

In this section, we present the details of the calculations to derive the effective
Hamiltonian around $\bm{K}_{D}=\bm{K}_{0} + \bm{A}$. Also, these calculations can be taken as 
an alternative proof of Eq.~(\ref{VP}), as pointed out below.   

Considering momenta close to the Dirac point $\bm{K}_{D}$ , 
i.e. $\bm{k}=\bm{K}_{D}+\bm{q}$, and expanding to first order in $\bm{q}$ and $\{\Delta_{n}\}$,
Hamiltonian (\ref{H}) transforms as
\begin{eqnarray}\label{G}
H = &-&\sum_{n=1}^{3}t_{n}
\begin{pmatrix}
0 & e^{-i(\bm{K}_{0} + \bm{A} +\bm{q})\cdot\bm{\delta}_{n}}\\
e^{i(\bm{K}_{0} + \bm{A} +\bm{q})\cdot\bm{\delta}_{n}} & 0
\end{pmatrix},\nonumber\\
 = &-&\sum_{n=1}^{3}t_{0}(1+\Delta_{n})
\begin{pmatrix}
0 & e^{-i\bm{K}_{0}\cdot\bm{\delta}_{n}}\\
e^{i\bm{K}_{0}\cdot\bm{\delta}_{n}} & 0
\end{pmatrix}\nonumber\\
&\times&(\bar{\bm{I}}+i\sigma_{z}\bm{A}\cdot\bm{\delta}_{n})(\bar{\bm{I}}+i\sigma_{z}\bm{q}\cdot\bm{\delta}_{n}),\nonumber\\
= &-&t_{0}\sum_{n=1}^{3}
(i\frac{\bm{\sigma}\cdot\bm{\delta}_{n}}{a}\sigma_{z}) \Bigl( \bar{\bm{I}} + i\sigma_{z}\bm{q}\cdot\bm{\delta}_{n}
+i\sigma_{z}\bm{A}\cdot\bm{\delta}_{n} \nonumber\\
&+& \Delta_{n}\bar{\bm{I}} -(\bm{q}\cdot\bm{\delta}_{n})(\bm{A}\cdot\bm{\delta}_{n})\bar{\bm{I}}
\Bigr),\label{EH}
\end{eqnarray}
where it has been assumed that the $\bm{A}$ vector is given by Eq.~(\ref{VP}) and $\bm{K}_{0}=(\frac{4\pi}{3\sqrt{3}a},0)$ has valley index $\xi=+1$. 
Collecting the contribution of each term in this expression,
one obtain:
\begin{equation}\label{T1}
-t_{0}\sum_{n=1}^{3}
(i\frac{\bm{\sigma}\cdot\bm{\delta}_{n}}{a}\sigma_{z})=0,
\end{equation}

\begin{equation}\label{T2}
-t_{0}\sum_{n=1}^{3}
(i\frac{\bm{\sigma}\cdot\bm{\delta}_{n}}{a}\sigma_{z})
(i\sigma_{z}\bm{q}\cdot\bm{\delta}_{n})=\hbar v_{F}\bm{\sigma}\cdot\bm{q},
\end{equation}

\begin{equation}\label{T3}
-t_{0}\sum_{n=1}^{3}
(i\frac{\bm{\sigma}\cdot\bm{\delta}_{n}}{a}\sigma_{z})
(i\sigma_{z}\bm{A}\cdot\bm{\delta}_{n})=\hbar v_{F}\bm{\sigma}\cdot\bm{A},
\end{equation}

\begin{equation}
-t_{0}\sum_{n=1}^{3}
(i\frac{\bm{\sigma}\cdot\bm{\delta}_{n}}{a}\sigma_{z})
\Delta_{n}=-\hbar v_{F}\bm{\sigma}\cdot\bm{A},
\end{equation}

\begin{equation}
t_{0}\sum_{n=1}^{3}
(i\frac{\bm{\sigma}\cdot\bm{\delta}_{n}}{a}\sigma_{z})
(\bm{q}\cdot\bm{\delta}_{n})(\bm{A}\cdot\bm{\delta}_{n})=\hbar v_{F}\bm{\sigma}\cdot\bar{\bm{\Delta}}\cdot\bm{q},
\end{equation}
where the $\bar{\bm{\Delta}}$ matrix is given by Eq. (\ref{Delta}). 
Then, taking into account the contribution of each term in
Eq.~(\ref{EH}), the effective
Dirac Hamiltonian around $K_{D}$ has the form 
\begin{equation}\label{DH_A}
H=\hbar
v_{F}\bm{\sigma}\cdot(\bar{\bm{I}}+\bar{\bm{\Delta}})\cdot\bm{q}.
\end{equation}

This result also proves that the Dirac point $\bm{K}_{D}$ is given by Eqs.(\ref{KD}-\ref{VP}). Note that in the Hamiltonian (\ref{DH_A}),
all terms are $\mathcal{O}(\bm{q})$, which is a consequence of an expansion around the real Dirac point and, therefore, this proves that 
the expression (\ref{VP}) for the shift $\bm{A}$ of the Dirac point is correct.

For $\bm{K}_{0}$ with valley index $\xi=-1$, the calculation is analogous, and the effective Dirac Hamiltonian results
\begin{equation}
H=\hbar
v_{F}\bm{\sigma}^{*}\cdot(\bar{\bm{I}}+\bar{\bm{\Delta}})\cdot\bm{q}.
\end{equation}
where $\bm{\sigma}^{*}=(\sigma_{x},-\sigma_{y})$.

\bibliography{biblioStrainedGraphene}

\begin{thebibliography}{56}%
\makeatletter
\providecommand \@ifxundefined [1]{%
 \@ifx{#1\undefined}
}%
\providecommand \@ifnum [1]{%
 \ifnum #1\expandafter \@firstoftwo
 \else \expandafter \@secondoftwo
 \fi
}%
\providecommand \@ifx [1]{%
 \ifx #1\expandafter \@firstoftwo
 \else \expandafter \@secondoftwo
 \fi
}%
\providecommand \natexlab [1]{#1}%
\providecommand \enquote  [1]{``#1''}%
\providecommand \bibnamefont  [1]{#1}%
\providecommand \bibfnamefont [1]{#1}%
\providecommand \citenamefont [1]{#1}%
\providecommand \href@noop [0]{\@secondoftwo}%
\providecommand \href [0]{\begingroup \@sanitize@url \@href}%
\providecommand \@href[1]{\@@startlink{#1}\@@href}%
\providecommand \@@href[1]{\endgroup#1\@@endlink}%
\providecommand \@sanitize@url [0]{\catcode `\\12\catcode `\$12\catcode
  `\&12\catcode `\#12\catcode `\^12\catcode `\_12\catcode `\%12\relax}%
\providecommand \@@startlink[1]{}%
\providecommand \@@endlink[0]{}%
\providecommand \url  [0]{\begingroup\@sanitize@url \@url }%
\providecommand \@url [1]{\endgroup\@href {#1}{\urlprefix }}%
\providecommand \urlprefix  [0]{URL }%
\providecommand \Eprint [0]{\href }%
\providecommand \doibase [0]{http://dx.doi.org/}%
\providecommand \selectlanguage [0]{\@gobble}%
\providecommand \bibinfo  [0]{\@secondoftwo}%
\providecommand \bibfield  [0]{\@secondoftwo}%
\providecommand \translation [1]{[#1]}%
\providecommand \BibitemOpen [0]{}%
\providecommand \bibitemStop [0]{}%
\providecommand \bibitemNoStop [0]{.\EOS\space}%
\providecommand \EOS [0]{\spacefactor3000\relax}%
\providecommand \BibitemShut  [1]{\csname bibitem#1\endcsname}%
\let\auto@bib@innerbib\@empty
\bibitem [{\citenamefont {Castro~Neto}\ \emph {et~al.}(2009)\citenamefont
  {Castro~Neto}, \citenamefont {Guinea}, \citenamefont {Peres}, \citenamefont
  {Novoselov},\ and\ \citenamefont {Geim}}]{Neto09}%
  \BibitemOpen
  \bibfield  {author} {\bibinfo {author} {\bibfnamefont {A.~H.}\ \bibnamefont
  {Castro~Neto}}, \bibinfo {author} {\bibfnamefont {F.}~\bibnamefont {Guinea}},
  \bibinfo {author} {\bibfnamefont {N.~M.~R.}\ \bibnamefont {Peres}}, \bibinfo
  {author} {\bibfnamefont {K.~S.}\ \bibnamefont {Novoselov}}, \ and\ \bibinfo
  {author} {\bibfnamefont {A.~K.}\ \bibnamefont {Geim}},\ }\href {\doibase
  10.1103/RevModPhys.81.109} {\bibfield  {journal} {\bibinfo  {journal} {Rev.
  Mod. Phys.}\ }\textbf {\bibinfo {volume} {81}},\ \bibinfo {pages} {109}
  (\bibinfo {year} {2009})}\BibitemShut {NoStop}%
\bibitem [{\citenamefont {Katsnelson}\ \emph {et~al.}(2006)\citenamefont
  {Katsnelson}, \citenamefont {Novoselov},\ and\ \citenamefont
  {Geim}}]{Katsnelson06}%
  \BibitemOpen
  \bibfield  {author} {\bibinfo {author} {\bibfnamefont {M.~I.}\ \bibnamefont
  {Katsnelson}}, \bibinfo {author} {\bibfnamefont {K.~S.}\ \bibnamefont
  {Novoselov}}, \ and\ \bibinfo {author} {\bibfnamefont {A.~K.}\ \bibnamefont
  {Geim}},\ }\href {\doibase 10.1038/nphys384} {\bibfield  {journal} {\bibinfo
  {journal} {Nat. Phys.}\ }\textbf {\bibinfo {volume} {2}},\ \bibinfo {pages}
  {620} (\bibinfo {year} {2006})}\BibitemShut {NoStop}%
\bibitem [{\citenamefont {Young}\ and\ \citenamefont {Kim}(2009)}]{Kim09}%
  \BibitemOpen
  \bibfield  {author} {\bibinfo {author} {\bibfnamefont {A.~F.}\ \bibnamefont
  {Young}}\ and\ \bibinfo {author} {\bibfnamefont {P.}~\bibnamefont {Kim}},\
  }\href {\doibase 10.1038/nphys1198} {\bibfield  {journal} {\bibinfo
  {journal} {Nat. Phys.}\ }\textbf {\bibinfo {volume} {5}},\ \bibinfo {pages}
  {222} (\bibinfo {year} {2009})}\BibitemShut {NoStop}%
\bibitem [{\citenamefont {Calogeracos}\ and\ \citenamefont
  {Dombey}(1999)}]{Calogeracos}%
  \BibitemOpen
  \bibfield  {author} {\bibinfo {author} {\bibfnamefont {A.}~\bibnamefont
  {Calogeracos}}\ and\ \bibinfo {author} {\bibfnamefont {N.}~\bibnamefont
  {Dombey}},\ }\href {\doibase 10.1080/001075199181387} {\bibfield  {journal}
  {\bibinfo  {journal} {Contemporary Physics}\ }\textbf {\bibinfo {volume}
  {40}},\ \bibinfo {pages} {313} (\bibinfo {year} {1999})}\BibitemShut
  {NoStop}%
\bibitem [{\citenamefont {Lee}\ \emph {et~al.}(2008)\citenamefont {Lee},
  \citenamefont {Wei}, \citenamefont {Kysar},\ and\ \citenamefont
  {Hone}}]{Lee08}%
  \BibitemOpen
  \bibfield  {author} {\bibinfo {author} {\bibfnamefont {C.}~\bibnamefont
  {Lee}}, \bibinfo {author} {\bibfnamefont {X.}~\bibnamefont {Wei}}, \bibinfo
  {author} {\bibfnamefont {J.~W.}\ \bibnamefont {Kysar}}, \ and\ \bibinfo
  {author} {\bibfnamefont {J.}~\bibnamefont {Hone}},\ }\href {\doibase
  10.1126/science.1157996} {\bibfield  {journal} {\bibinfo  {journal}
  {Science}\ }\textbf {\bibinfo {volume} {321}},\ \bibinfo {pages} {385}
  (\bibinfo {year} {2008})}\BibitemShut {NoStop}%
\bibitem [{\citenamefont {Castellanos-Gomez}\ \emph {et~al.}(2015)\citenamefont
  {Castellanos-Gomez}, \citenamefont {Singh}, \citenamefont {van~der Zant},\
  and\ \citenamefont {Steele}}]{Castellanos}%
  \BibitemOpen
  \bibfield  {author} {\bibinfo {author} {\bibfnamefont {A.}~\bibnamefont
  {Castellanos-Gomez}}, \bibinfo {author} {\bibfnamefont {V.}~\bibnamefont
  {Singh}}, \bibinfo {author} {\bibfnamefont {H.~S.~J.}\ \bibnamefont {van~der
  Zant}}, \ and\ \bibinfo {author} {\bibfnamefont {G.~A.}\ \bibnamefont
  {Steele}},\ }\href {\doibase 10.1002/andp.201400153} {\bibfield  {journal}
  {\bibinfo  {journal} {Annalen der Physik}\ }\textbf {\bibinfo {volume}
  {527}},\ \bibinfo {pages} {27} (\bibinfo {year} {2015})}\BibitemShut
  {NoStop}%
\bibitem [{\citenamefont {Pereira}\ and\ \citenamefont
  {Castro~Neto}(2009)}]{Pereira09b}%
  \BibitemOpen
  \bibfield  {author} {\bibinfo {author} {\bibfnamefont {V.~M.}\ \bibnamefont
  {Pereira}}\ and\ \bibinfo {author} {\bibfnamefont {A.~H.}\ \bibnamefont
  {Castro~Neto}},\ }\href {\doibase 10.1103/PhysRevLett.103.046801} {\bibfield
  {journal} {\bibinfo  {journal} {Phys. Rev. Lett.}\ }\textbf {\bibinfo
  {volume} {103}},\ \bibinfo {pages} {046801} (\bibinfo {year}
  {2009})}\BibitemShut {NoStop}%
\bibitem [{\citenamefont {Guinea}(2012)}]{Guinea12}%
  \BibitemOpen
  \bibfield  {author} {\bibinfo {author} {\bibfnamefont {F.}~\bibnamefont
  {Guinea}},\ }\href {\doibase http://dx.doi.org/10.1016/j.ssc.2012.04.019}
  {\bibfield  {journal} {\bibinfo  {journal} {Solid State Communications}\
  }\textbf {\bibinfo {volume} {152}},\ \bibinfo {pages} {1437 } (\bibinfo
  {year} {2012})}\BibitemShut {NoStop}%
\bibitem [{\citenamefont {Wang}\ \emph {et~al.}(2015)\citenamefont {Wang},
  \citenamefont {Wang},\ and\ \citenamefont {Liu}}]{Wang15}%
  \BibitemOpen
  \bibfield  {author} {\bibinfo {author} {\bibfnamefont {B.}~\bibnamefont
  {Wang}}, \bibinfo {author} {\bibfnamefont {Y.}~\bibnamefont {Wang}}, \ and\
  \bibinfo {author} {\bibfnamefont {Y.}~\bibnamefont {Liu}},\ }\href {\doibase
  10.1142/S1793604715300017} {\bibfield  {journal} {\bibinfo  {journal}
  {Functional Materials Letters}\ }\textbf {\bibinfo {volume} {08}},\ \bibinfo
  {pages} {1530001} (\bibinfo {year} {2015})}\BibitemShut {NoStop}%
\bibitem [{\citenamefont {Amorim}\ \emph {et~al.}(2015)\citenamefont {Amorim},
  \citenamefont {Cortijo}, \citenamefont {de~Juan}, \citenamefont {Grushin},
  \citenamefont {Guinea}, \citenamefont {Guti\'errez-Rubio}, \citenamefont
  {Ochoa}, \citenamefont {Parente}, \citenamefont {Rold\'an}, \citenamefont
  {San-Jos\'e}, \citenamefont {Schiefele}, \citenamefont {Sturla},\ and\
  \citenamefont {Vozmediano}}]{Amorim}%
  \BibitemOpen
  \bibfield  {author} {\bibinfo {author} {\bibfnamefont {B.}~\bibnamefont
  {Amorim}}, \bibinfo {author} {\bibfnamefont {A.}~\bibnamefont {Cortijo}},
  \bibinfo {author} {\bibfnamefont {F.}~\bibnamefont {de~Juan}}, \bibinfo
  {author} {\bibfnamefont {A.~G.}\ \bibnamefont {Grushin}}, \bibinfo {author}
  {\bibfnamefont {F.}~\bibnamefont {Guinea}}, \bibinfo {author} {\bibfnamefont
  {A.}~\bibnamefont {Guti\'errez-Rubio}}, \bibinfo {author} {\bibfnamefont
  {H.}~\bibnamefont {Ochoa}}, \bibinfo {author} {\bibfnamefont
  {V.}~\bibnamefont {Parente}}, \bibinfo {author} {\bibfnamefont
  {R.}~\bibnamefont {Rold\'an}}, \bibinfo {author} {\bibfnamefont
  {P.}~\bibnamefont {San-Jos\'e}}, \bibinfo {author} {\bibfnamefont
  {J.}~\bibnamefont {Schiefele}}, \bibinfo {author} {\bibfnamefont
  {M.}~\bibnamefont {Sturla}}, \ and\ \bibinfo {author} {\bibfnamefont
  {M.~A.~H.}\ \bibnamefont {Vozmediano}},\ }\href
  {http://arxiv.org/abs/1503.00747} {\bibfield  {journal} {\bibinfo  {journal}
  {arXiv:1503.00747}\ } (\bibinfo {year} {2015})}\BibitemShut {NoStop}%
\bibitem [{\citenamefont {Pereira}\ \emph {et~al.}(2009)\citenamefont
  {Pereira}, \citenamefont {Castro~Neto},\ and\ \citenamefont
  {Peres}}]{Pereira09a}%
  \BibitemOpen
  \bibfield  {author} {\bibinfo {author} {\bibfnamefont {V.~M.}\ \bibnamefont
  {Pereira}}, \bibinfo {author} {\bibfnamefont {A.~H.}\ \bibnamefont
  {Castro~Neto}}, \ and\ \bibinfo {author} {\bibfnamefont {N.~M.~R.}\
  \bibnamefont {Peres}},\ }\href {\doibase 10.1103/PhysRevB.80.045401}
  {\bibfield  {journal} {\bibinfo  {journal} {Phys. Rev. B}\ }\textbf {\bibinfo
  {volume} {80}},\ \bibinfo {pages} {045401} (\bibinfo {year}
  {2009})}\BibitemShut {NoStop}%
\bibitem [{\citenamefont {Li}\ \emph {et~al.}(2010)\citenamefont {Li},
  \citenamefont {Jiang}, \citenamefont {Liu},\ and\ \citenamefont
  {Liu}}]{YangLi10}%
  \BibitemOpen
  \bibfield  {author} {\bibinfo {author} {\bibfnamefont {Y.}~\bibnamefont
  {Li}}, \bibinfo {author} {\bibfnamefont {X.}~\bibnamefont {Jiang}}, \bibinfo
  {author} {\bibfnamefont {Z.}~\bibnamefont {Liu}}, \ and\ \bibinfo {author}
  {\bibfnamefont {Z.}~\bibnamefont {Liu}},\ }\href {\doibase
  10.1007/s12274-010-0015-7} {\bibfield  {journal} {\bibinfo  {journal} {Nano
  Research}\ }\textbf {\bibinfo {volume} {3}},\ \bibinfo {pages} {545}
  (\bibinfo {year} {2010})}\BibitemShut {NoStop}%
\bibitem [{\citenamefont {Cocco}\ \emph {et~al.}(2010)\citenamefont {Cocco},
  \citenamefont {Cadelano},\ and\ \citenamefont {Colombo}}]{Colombo}%
  \BibitemOpen
  \bibfield  {author} {\bibinfo {author} {\bibfnamefont {G.}~\bibnamefont
  {Cocco}}, \bibinfo {author} {\bibfnamefont {E.}~\bibnamefont {Cadelano}}, \
  and\ \bibinfo {author} {\bibfnamefont {L.}~\bibnamefont {Colombo}},\ }\href
  {\doibase 10.1103/PhysRevB.81.241412} {\bibfield  {journal} {\bibinfo
  {journal} {Phys. Rev. B}\ }\textbf {\bibinfo {volume} {81}},\ \bibinfo
  {pages} {241412} (\bibinfo {year} {2010})}\BibitemShut {NoStop}%
\bibitem [{\citenamefont {Gui}\ \emph {et~al.}(2015)\citenamefont {Gui},
  \citenamefont {Morgan}, \citenamefont {Booske}, \citenamefont {Zhong},\ and\
  \citenamefont {Ma}}]{Gui15}%
  \BibitemOpen
  \bibfield  {author} {\bibinfo {author} {\bibfnamefont {G.}~\bibnamefont
  {Gui}}, \bibinfo {author} {\bibfnamefont {D.}~\bibnamefont {Morgan}},
  \bibinfo {author} {\bibfnamefont {J.}~\bibnamefont {Booske}}, \bibinfo
  {author} {\bibfnamefont {J.}~\bibnamefont {Zhong}}, \ and\ \bibinfo {author}
  {\bibfnamefont {Z.}~\bibnamefont {Ma}},\ }\href {\doibase
  http://dx.doi.org/10.1063/1.4907410} {\bibfield  {journal} {\bibinfo
  {journal} {Applied Physics Letters}\ }\textbf {\bibinfo {volume} {106}},\
  \bibinfo {eid} {053113} (\bibinfo {year} {2015})}\BibitemShut {NoStop}%
\bibitem [{\citenamefont {Levy}\ \emph {et~al.}(2010)\citenamefont {Levy},
  \citenamefont {Burke}, \citenamefont {Meaker}, \citenamefont {Panlasigui},
  \citenamefont {Zettl}, \citenamefont {Guinea}, \citenamefont {Neto},\ and\
  \citenamefont {Crommie}}]{Levy}%
  \BibitemOpen
  \bibfield  {author} {\bibinfo {author} {\bibfnamefont {N.}~\bibnamefont
  {Levy}}, \bibinfo {author} {\bibfnamefont {S.~A.}\ \bibnamefont {Burke}},
  \bibinfo {author} {\bibfnamefont {K.~L.}\ \bibnamefont {Meaker}}, \bibinfo
  {author} {\bibfnamefont {M.}~\bibnamefont {Panlasigui}}, \bibinfo {author}
  {\bibfnamefont {A.}~\bibnamefont {Zettl}}, \bibinfo {author} {\bibfnamefont
  {F.}~\bibnamefont {Guinea}}, \bibinfo {author} {\bibfnamefont {A.~H.~C.}\
  \bibnamefont {Neto}}, \ and\ \bibinfo {author} {\bibfnamefont {M.~F.}\
  \bibnamefont {Crommie}},\ }\href {\doibase 10.1126/science.1191700}
  {\bibfield  {journal} {\bibinfo  {journal} {Science}\ }\textbf {\bibinfo
  {volume} {329}},\ \bibinfo {pages} {544} (\bibinfo {year}
  {2010})}\BibitemShut {NoStop}%
\bibitem [{\citenamefont {Lu}\ \emph {et~al.}(2012)\citenamefont {Lu},
  \citenamefont {Neto},\ and\ \citenamefont {Loh}}]{Jiong}%
  \BibitemOpen
  \bibfield  {author} {\bibinfo {author} {\bibfnamefont {J.}~\bibnamefont
  {Lu}}, \bibinfo {author} {\bibfnamefont {A.~C.}\ \bibnamefont {Neto}}, \ and\
  \bibinfo {author} {\bibfnamefont {K.~P.}\ \bibnamefont {Loh}},\ }\href
  {\doibase 10.1038/ncomms1818} {\bibfield  {journal} {\bibinfo  {journal} {Nat
  Commun}\ }\textbf {\bibinfo {volume} {3}},\ \bibinfo {pages} {823} (\bibinfo
  {year} {2012})}\BibitemShut {NoStop}%
\bibitem [{\citenamefont {Suzuura}\ and\ \citenamefont {Ando}(2002)}]{Ando02}%
  \BibitemOpen
  \bibfield  {author} {\bibinfo {author} {\bibfnamefont {H.}~\bibnamefont
  {Suzuura}}\ and\ \bibinfo {author} {\bibfnamefont {T.}~\bibnamefont {Ando}},\
  }\href {\doibase 10.1103/PhysRevB.65.235412} {\bibfield  {journal} {\bibinfo
  {journal} {Phys. Rev. B}\ }\textbf {\bibinfo {volume} {65}},\ \bibinfo
  {pages} {235412} (\bibinfo {year} {2002})}\BibitemShut {NoStop}%
\bibitem [{\citenamefont {Morpurgo}\ and\ \citenamefont
  {Guinea}(2006)}]{Morpurgo}%
  \BibitemOpen
  \bibfield  {author} {\bibinfo {author} {\bibfnamefont {A.~F.}\ \bibnamefont
  {Morpurgo}}\ and\ \bibinfo {author} {\bibfnamefont {F.}~\bibnamefont
  {Guinea}},\ }\href {\doibase 10.1103/PhysRevLett.97.196804} {\bibfield
  {journal} {\bibinfo  {journal} {Phys. Rev. Lett.}\ }\textbf {\bibinfo
  {volume} {97}},\ \bibinfo {pages} {196804} (\bibinfo {year}
  {2006})}\BibitemShut {NoStop}%
\bibitem [{\citenamefont {Morozov}\ \emph {et~al.}(2006)\citenamefont
  {Morozov}, \citenamefont {Novoselov}, \citenamefont {Katsnelson},
  \citenamefont {Schedin}, \citenamefont {Ponomarenko}, \citenamefont {Jiang},\
  and\ \citenamefont {Geim}}]{Morozov}%
  \BibitemOpen
  \bibfield  {author} {\bibinfo {author} {\bibfnamefont {S.~V.}\ \bibnamefont
  {Morozov}}, \bibinfo {author} {\bibfnamefont {K.~S.}\ \bibnamefont
  {Novoselov}}, \bibinfo {author} {\bibfnamefont {M.~I.}\ \bibnamefont
  {Katsnelson}}, \bibinfo {author} {\bibfnamefont {F.}~\bibnamefont {Schedin}},
  \bibinfo {author} {\bibfnamefont {L.~A.}\ \bibnamefont {Ponomarenko}},
  \bibinfo {author} {\bibfnamefont {D.}~\bibnamefont {Jiang}}, \ and\ \bibinfo
  {author} {\bibfnamefont {A.~K.}\ \bibnamefont {Geim}},\ }\href {\doibase
  10.1103/PhysRevLett.97.016801} {\bibfield  {journal} {\bibinfo  {journal}
  {Phys. Rev. Lett.}\ }\textbf {\bibinfo {volume} {97}},\ \bibinfo {pages}
  {016801} (\bibinfo {year} {2006})}\BibitemShut {NoStop}%
\bibitem [{\citenamefont {Guinea}\ \emph {et~al.}(2010)\citenamefont {Guinea},
  \citenamefont {Katsnelson},\ and\ \citenamefont {Geim}}]{Guinea10a}%
  \BibitemOpen
  \bibfield  {author} {\bibinfo {author} {\bibfnamefont {F.}~\bibnamefont
  {Guinea}}, \bibinfo {author} {\bibfnamefont {M.~I.}\ \bibnamefont
  {Katsnelson}}, \ and\ \bibinfo {author} {\bibfnamefont {A.~K.}\ \bibnamefont
  {Geim}},\ }\href {\doibase 10.1038/nphys1420} {\bibfield  {journal} {\bibinfo
   {journal} {Nat Phys}\ }\textbf {\bibinfo {volume} {6}},\ \bibinfo {pages}
  {30} (\bibinfo {year} {2010})}\BibitemShut {NoStop}%
\bibitem [{\citenamefont {Sloan}\ \emph {et~al.}(2013)\citenamefont {Sloan},
  \citenamefont {Sanjuan}, \citenamefont {Wang}, \citenamefont {Horvath},\ and\
  \citenamefont {Barraza-Lopez}}]{Salvador13}%
  \BibitemOpen
  \bibfield  {author} {\bibinfo {author} {\bibfnamefont {J.~V.}\ \bibnamefont
  {Sloan}}, \bibinfo {author} {\bibfnamefont {A.~A.~P.}\ \bibnamefont
  {Sanjuan}}, \bibinfo {author} {\bibfnamefont {Z.}~\bibnamefont {Wang}},
  \bibinfo {author} {\bibfnamefont {C.}~\bibnamefont {Horvath}}, \ and\
  \bibinfo {author} {\bibfnamefont {S.}~\bibnamefont {Barraza-Lopez}},\ }\href
  {\doibase 10.1103/PhysRevB.87.155436} {\bibfield  {journal} {\bibinfo
  {journal} {Phys. Rev. B}\ }\textbf {\bibinfo {volume} {87}},\ \bibinfo
  {pages} {155436} (\bibinfo {year} {2013})}\BibitemShut {NoStop}%
\bibitem [{\citenamefont {Gradinar}\ \emph {et~al.}(2013)\citenamefont
  {Gradinar}, \citenamefont {Mucha-Kruczy\ifmmode~\acute{n}\else
  \'{n}\fi{}ski}, \citenamefont {Schomerus},\ and\ \citenamefont
  {Fal'ko}}]{Falko13}%
  \BibitemOpen
  \bibfield  {author} {\bibinfo {author} {\bibfnamefont {D.~A.}\ \bibnamefont
  {Gradinar}}, \bibinfo {author} {\bibfnamefont {M.}~\bibnamefont
  {Mucha-Kruczy\ifmmode~\acute{n}\else \'{n}\fi{}ski}}, \bibinfo {author}
  {\bibfnamefont {H.}~\bibnamefont {Schomerus}}, \ and\ \bibinfo {author}
  {\bibfnamefont {V.~I.}\ \bibnamefont {Fal'ko}},\ }\href {\doibase
  10.1103/PhysRevLett.110.266801} {\bibfield  {journal} {\bibinfo  {journal}
  {Phys. Rev. Lett.}\ }\textbf {\bibinfo {volume} {110}},\ \bibinfo {pages}
  {266801} (\bibinfo {year} {2013})}\BibitemShut {NoStop}%
\bibitem [{\citenamefont {He}\ and\ \citenamefont {He}(2013)}]{LinHe13}%
  \BibitemOpen
  \bibfield  {author} {\bibinfo {author} {\bibfnamefont {W.-Y.}\ \bibnamefont
  {He}}\ and\ \bibinfo {author} {\bibfnamefont {L.}~\bibnamefont {He}},\ }\href
  {\doibase 10.1103/PhysRevB.88.085411} {\bibfield  {journal} {\bibinfo
  {journal} {Phys. Rev. B}\ }\textbf {\bibinfo {volume} {88}},\ \bibinfo
  {pages} {085411} (\bibinfo {year} {2013})}\BibitemShut {NoStop}%
\bibitem [{\citenamefont {Carrillo-Bastos}\ \emph {et~al.}(2014)\citenamefont
  {Carrillo-Bastos}, \citenamefont {Faria}, \citenamefont {Latg\'e},
  \citenamefont {Mireles},\ and\ \citenamefont {Sandler}}]{Sandler14}%
  \BibitemOpen
  \bibfield  {author} {\bibinfo {author} {\bibfnamefont {R.}~\bibnamefont
  {Carrillo-Bastos}}, \bibinfo {author} {\bibfnamefont {D.}~\bibnamefont
  {Faria}}, \bibinfo {author} {\bibfnamefont {A.}~\bibnamefont {Latg\'e}},
  \bibinfo {author} {\bibfnamefont {F.}~\bibnamefont {Mireles}}, \ and\
  \bibinfo {author} {\bibfnamefont {N.}~\bibnamefont {Sandler}},\ }\href
  {\doibase 10.1103/PhysRevB.90.041411} {\bibfield  {journal} {\bibinfo
  {journal} {Phys. Rev. B}\ }\textbf {\bibinfo {volume} {90}},\ \bibinfo
  {pages} {041411} (\bibinfo {year} {2014})}\BibitemShut {NoStop}%
\bibitem [{\citenamefont {Qi}\ \emph {et~al.}(2014)\citenamefont {Qi},
  \citenamefont {Kitt}, \citenamefont {Park}, \citenamefont {Pereira},
  \citenamefont {Campbell},\ and\ \citenamefont {Castro~Neto}}]{Zenan14}%
  \BibitemOpen
  \bibfield  {author} {\bibinfo {author} {\bibfnamefont {Z.}~\bibnamefont
  {Qi}}, \bibinfo {author} {\bibfnamefont {A.~L.}\ \bibnamefont {Kitt}},
  \bibinfo {author} {\bibfnamefont {H.~S.}\ \bibnamefont {Park}}, \bibinfo
  {author} {\bibfnamefont {V.~M.}\ \bibnamefont {Pereira}}, \bibinfo {author}
  {\bibfnamefont {D.~K.}\ \bibnamefont {Campbell}}, \ and\ \bibinfo {author}
  {\bibfnamefont {A.~H.}\ \bibnamefont {Castro~Neto}},\ }\href {\doibase
  10.1103/PhysRevB.90.125419} {\bibfield  {journal} {\bibinfo  {journal} {Phys.
  Rev. B}\ }\textbf {\bibinfo {volume} {90}},\ \bibinfo {pages} {125419}
  (\bibinfo {year} {2014})}\BibitemShut {NoStop}%
\bibitem [{\citenamefont {Burgos}\ \emph {et~al.}(2015)\citenamefont {Burgos},
  \citenamefont {Warnes}, \citenamefont {Lima},\ and\ \citenamefont
  {Lewenkopf}}]{Burgos15}%
  \BibitemOpen
  \bibfield  {author} {\bibinfo {author} {\bibfnamefont {R.}~\bibnamefont
  {Burgos}}, \bibinfo {author} {\bibfnamefont {J.}~\bibnamefont {Warnes}},
  \bibinfo {author} {\bibfnamefont {L.~R.~F.}\ \bibnamefont {Lima}}, \ and\
  \bibinfo {author} {\bibfnamefont {C.}~\bibnamefont {Lewenkopf}},\ }\href
  {\doibase 10.1103/PhysRevB.91.115403} {\bibfield  {journal} {\bibinfo
  {journal} {Phys. Rev. B}\ }\textbf {\bibinfo {volume} {91}},\ \bibinfo
  {pages} {115403} (\bibinfo {year} {2015})}\BibitemShut {NoStop}%
\bibitem [{\citenamefont {Guassi}\ \emph {et~al.}(2015)\citenamefont {Guassi},
  \citenamefont {Diniz}, \citenamefont {Sandler},\ and\ \citenamefont
  {Qu}}]{Sandler15}%
  \BibitemOpen
  \bibfield  {author} {\bibinfo {author} {\bibfnamefont {M.~R.}\ \bibnamefont
  {Guassi}}, \bibinfo {author} {\bibfnamefont {G.~S.}\ \bibnamefont {Diniz}},
  \bibinfo {author} {\bibfnamefont {N.}~\bibnamefont {Sandler}}, \ and\
  \bibinfo {author} {\bibfnamefont {F.}~\bibnamefont {Qu}},\ }\href {\doibase
  10.1103/PhysRevB.92.075426} {\bibfield  {journal} {\bibinfo  {journal} {Phys.
  Rev. B}\ }\textbf {\bibinfo {volume} {92}},\ \bibinfo {pages} {075426}
  (\bibinfo {year} {2015})}\BibitemShut {NoStop}%
\bibitem [{\citenamefont {Gruji\ifmmode~\acute{c}\else \'{c}\fi{}}\ \emph
  {et~al.}(2014)\citenamefont {Gruji\ifmmode~\acute{c}\else \'{c}\fi{}},
  \citenamefont {Tadi\ifmmode~\acute{c}\else \'{c}\fi{}},\ and\ \citenamefont
  {Peeters}}]{Peeters15}%
  \BibitemOpen
  \bibfield  {author} {\bibinfo {author} {\bibfnamefont {M.~M.}\ \bibnamefont
  {Gruji\ifmmode~\acute{c}\else \'{c}\fi{}}}, \bibinfo {author} {\bibfnamefont
  {M.~Z.}\ \bibnamefont {Tadi\ifmmode~\acute{c}\else \'{c}\fi{}}}, \ and\
  \bibinfo {author} {\bibfnamefont {F.~M.}\ \bibnamefont {Peeters}},\ }\href
  {\doibase 10.1103/PhysRevLett.113.046601} {\bibfield  {journal} {\bibinfo
  {journal} {Phys. Rev. Lett.}\ }\textbf {\bibinfo {volume} {113}},\ \bibinfo
  {pages} {046601} (\bibinfo {year} {2014})}\BibitemShut {NoStop}%
\bibitem [{\citenamefont {Midtvedt}\ \emph {et~al.}()\citenamefont {Midtvedt},
  \citenamefont {Lewenkopf},\ and\ \citenamefont {Croy}}]{Croy15}%
  \BibitemOpen
  \bibfield  {author} {\bibinfo {author} {\bibfnamefont {D.}~\bibnamefont
  {Midtvedt}}, \bibinfo {author} {\bibfnamefont {C.~H.}\ \bibnamefont
  {Lewenkopf}}, \ and\ \bibinfo {author} {\bibfnamefont {A.}~\bibnamefont
  {Croy}},\ }\href {http://arxiv.org/abs/1509.02365} {\bibinfo  {journal}
  {arXiv:1509.02365}\ }\BibitemShut {NoStop}%
\bibitem [{\citenamefont {Rostami}\ \emph {et~al.}(2015)\citenamefont
  {Rostami}, \citenamefont {Rold\'an}, \citenamefont {Cappelluti},
  \citenamefont {Asgari},\ and\ \citenamefont {Guinea}}]{Rostami}%
  \BibitemOpen
\bibfield  {journal} {  }\bibfield  {author} {\bibinfo {author} {\bibfnamefont
  {H.}~\bibnamefont {Rostami}}, \bibinfo {author} {\bibfnamefont
  {R.}~\bibnamefont {Rold\'an}}, \bibinfo {author} {\bibfnamefont
  {E.}~\bibnamefont {Cappelluti}}, \bibinfo {author} {\bibfnamefont
  {R.}~\bibnamefont {Asgari}}, \ and\ \bibinfo {author} {\bibfnamefont
  {F.}~\bibnamefont {Guinea}},\ }\href {\doibase 10.1103/PhysRevB.92.195402}
  {\bibfield  {journal} {\bibinfo  {journal} {Phys. Rev. B}\ }\textbf {\bibinfo
  {volume} {92}},\ \bibinfo {pages} {195402} (\bibinfo {year}
  {2015})}\BibitemShut {NoStop}%
\bibitem [{\citenamefont {Cortijo}\ \emph {et~al.}(2015)\citenamefont
  {Cortijo}, \citenamefont {Ferreir\'os}, \citenamefont {Landsteiner},\ and\
  \citenamefont {Vozmediano}}]{Cortijo15}%
  \BibitemOpen
  \bibfield  {author} {\bibinfo {author} {\bibfnamefont {A.}~\bibnamefont
  {Cortijo}}, \bibinfo {author} {\bibfnamefont {Y.}~\bibnamefont
  {Ferreir\'os}}, \bibinfo {author} {\bibfnamefont {K.}~\bibnamefont
  {Landsteiner}}, \ and\ \bibinfo {author} {\bibfnamefont {M.~A.~H.}\
  \bibnamefont {Vozmediano}},\ }\href {\doibase 10.1103/PhysRevLett.115.177202}
  {\bibfield  {journal} {\bibinfo  {journal} {Phys. Rev. Lett.}\ }\textbf
  {\bibinfo {volume} {115}},\ \bibinfo {pages} {177202} (\bibinfo {year}
  {2015})}\BibitemShut {NoStop}%
\bibitem [{\citenamefont {Bae}\ \emph {et~al.}(2013)\citenamefont {Bae},
  \citenamefont {Lee}, \citenamefont {Sharma}, \citenamefont {Lee},
  \citenamefont {Kim},\ and\ \citenamefont {Ahn}}]{Bae13}%
  \BibitemOpen
  \bibfield  {author} {\bibinfo {author} {\bibfnamefont {S.-H.}\ \bibnamefont
  {Bae}}, \bibinfo {author} {\bibfnamefont {Y.}~\bibnamefont {Lee}}, \bibinfo
  {author} {\bibfnamefont {B.~K.}\ \bibnamefont {Sharma}}, \bibinfo {author}
  {\bibfnamefont {H.-J.}\ \bibnamefont {Lee}}, \bibinfo {author} {\bibfnamefont
  {J.-H.}\ \bibnamefont {Kim}}, \ and\ \bibinfo {author} {\bibfnamefont
  {J.-H.}\ \bibnamefont {Ahn}},\ }\href {\doibase
  http://dx.doi.org/10.1016/j.carbon.2012.08.048} {\bibfield  {journal}
  {\bibinfo  {journal} {Carbon}\ }\textbf {\bibinfo {volume} {51}},\ \bibinfo
  {pages} {236 } (\bibinfo {year} {2013})}\BibitemShut {NoStop}%
\bibitem [{\citenamefont {Shimano}\ \emph {et~al.}(2013)\citenamefont
  {Shimano}, \citenamefont {Yumoto}, \citenamefont {Yoo}, \citenamefont
  {Matsunaga}, \citenamefont {Tanabe}, \citenamefont {Hibino}, \citenamefont
  {Morimoto},\ and\ \citenamefont {Aoki}}]{Shimano}%
  \BibitemOpen
  \bibfield  {author} {\bibinfo {author} {\bibfnamefont {R.}~\bibnamefont
  {Shimano}}, \bibinfo {author} {\bibfnamefont {G.}~\bibnamefont {Yumoto}},
  \bibinfo {author} {\bibfnamefont {J.~Y.}\ \bibnamefont {Yoo}}, \bibinfo
  {author} {\bibfnamefont {R.}~\bibnamefont {Matsunaga}}, \bibinfo {author}
  {\bibfnamefont {S.}~\bibnamefont {Tanabe}}, \bibinfo {author} {\bibfnamefont
  {H.}~\bibnamefont {Hibino}}, \bibinfo {author} {\bibfnamefont
  {T.}~\bibnamefont {Morimoto}}, \ and\ \bibinfo {author} {\bibfnamefont
  {H.}~\bibnamefont {Aoki}},\ }\href {\doibase 10.1038/ncomms2866} {\bibfield
  {journal} {\bibinfo  {journal} {Nat Commun}\ }\textbf {\bibinfo {volume}
  {4}},\ \bibinfo {pages} {1841} (\bibinfo {year} {2013})}\BibitemShut
  {NoStop}%
\bibitem [{\citenamefont {Dong}\ \emph {et~al.}(2014)\citenamefont {Dong},
  \citenamefont {Wang}, \citenamefont {Liu}, \citenamefont {Chen},
  \citenamefont {Jiang}, \citenamefont {Xin}, \citenamefont {Xing},\ and\
  \citenamefont {Tian}}]{Bin14}%
  \BibitemOpen
  \bibfield  {author} {\bibinfo {author} {\bibfnamefont {B.}~\bibnamefont
  {Dong}}, \bibinfo {author} {\bibfnamefont {P.}~\bibnamefont {Wang}}, \bibinfo
  {author} {\bibfnamefont {Z.-B.}\ \bibnamefont {Liu}}, \bibinfo {author}
  {\bibfnamefont {X.-D.}\ \bibnamefont {Chen}}, \bibinfo {author}
  {\bibfnamefont {W.-S.}\ \bibnamefont {Jiang}}, \bibinfo {author}
  {\bibfnamefont {W.}~\bibnamefont {Xin}}, \bibinfo {author} {\bibfnamefont
  {F.}~\bibnamefont {Xing}}, \ and\ \bibinfo {author} {\bibfnamefont {J.-G.}\
  \bibnamefont {Tian}},\ }\href@noop {} {\bibfield  {journal} {\bibinfo
  {journal} {Nanotechnology}\ }\textbf {\bibinfo {volume} {25}},\ \bibinfo
  {pages} {455707} (\bibinfo {year} {2014})}\BibitemShut {NoStop}%
\bibitem [{\citenamefont {Oliva-Leyva}\ and\ \citenamefont
  {Naumis}(2013)}]{Our13}%
  \BibitemOpen
  \bibfield  {author} {\bibinfo {author} {\bibfnamefont {M.}~\bibnamefont
  {Oliva-Leyva}}\ and\ \bibinfo {author} {\bibfnamefont {G.~G.}\ \bibnamefont
  {Naumis}},\ }\href {\doibase 10.1103/PhysRevB.88.085430} {\bibfield
  {journal} {\bibinfo  {journal} {Phys. Rev. B}\ }\textbf {\bibinfo {volume}
  {88}},\ \bibinfo {pages} {085430} (\bibinfo {year} {2013})}\BibitemShut
  {NoStop}%
\bibitem [{\citenamefont {Pereira}\ \emph {et~al.}(2010)\citenamefont
  {Pereira}, \citenamefont {Ribeiro}, \citenamefont {Peres},\ and\
  \citenamefont {Castro~Neto}}]{Pereira10}%
  \BibitemOpen
  \bibfield  {author} {\bibinfo {author} {\bibfnamefont {V.~M.}\ \bibnamefont
  {Pereira}}, \bibinfo {author} {\bibfnamefont {R.~M.}\ \bibnamefont
  {Ribeiro}}, \bibinfo {author} {\bibfnamefont {N.~M.~R.}\ \bibnamefont
  {Peres}}, \ and\ \bibinfo {author} {\bibfnamefont {A.~H.}\ \bibnamefont
  {Castro~Neto}},\ }\href {http://stacks.iop.org/0295-5075/92/i=6/a=67001}
  {\bibfield  {journal} {\bibinfo  {journal} {EPL}\ }\textbf {\bibinfo {volume}
  {92}},\ \bibinfo {pages} {67001} (\bibinfo {year} {2010})}\BibitemShut
  {NoStop}%
\bibitem [{\citenamefont {Ni}\ \emph {et~al.}(2014)\citenamefont {Ni},
  \citenamefont {Yang}, \citenamefont {Ji}, \citenamefont {Baeck},
  \citenamefont {Toh}, \citenamefont {Ahn}, \citenamefont {Pereira},\ and\
  \citenamefont {\"Ozyilmaz}}]{Pereira14}%
  \BibitemOpen
  \bibfield  {author} {\bibinfo {author} {\bibfnamefont {G.-X.}\ \bibnamefont
  {Ni}}, \bibinfo {author} {\bibfnamefont {H.-Z.}\ \bibnamefont {Yang}},
  \bibinfo {author} {\bibfnamefont {W.}~\bibnamefont {Ji}}, \bibinfo {author}
  {\bibfnamefont {S.-J.}\ \bibnamefont {Baeck}}, \bibinfo {author}
  {\bibfnamefont {C.-T.}\ \bibnamefont {Toh}}, \bibinfo {author} {\bibfnamefont
  {J.-H.}\ \bibnamefont {Ahn}}, \bibinfo {author} {\bibfnamefont {V.~M.}\
  \bibnamefont {Pereira}}, \ and\ \bibinfo {author} {\bibfnamefont
  {B.}~\bibnamefont {\"Ozyilmaz}},\ }\href {\doibase 10.1002/adma.201304156}
  {\bibfield  {journal} {\bibinfo  {journal} {Advanced Materials}\ }\textbf
  {\bibinfo {volume} {26}},\ \bibinfo {pages} {1081} (\bibinfo {year}
  {2014})}\BibitemShut {NoStop}%
\bibitem [{\citenamefont {Oliva-Leyva}\ and\ \citenamefont
  {Naumis}(2015{\natexlab{a}})}]{Our15a}%
  \BibitemOpen
  \bibfield  {author} {\bibinfo {author} {\bibfnamefont {M.}~\bibnamefont
  {Oliva-Leyva}}\ and\ \bibinfo {author} {\bibfnamefont {G.~G.}\ \bibnamefont
  {Naumis}},\ }\href {http://stacks.iop.org/2053-1583/2/i=2/a=025001}
  {\bibfield  {journal} {\bibinfo  {journal} {2D Materials}\ }\textbf {\bibinfo
  {volume} {2}},\ \bibinfo {pages} {025001} (\bibinfo {year}
  {2015}{\natexlab{a}})}\BibitemShut {NoStop}%
\bibitem [{\citenamefont {Tarruell}\ \emph {et~al.}(2012)\citenamefont
  {Tarruell}, \citenamefont {Greif}, \citenamefont {Uehlinger}, \citenamefont
  {Jotzu},\ and\ \citenamefont {Esslinger}}]{Tarruell}%
  \BibitemOpen
  \bibfield  {author} {\bibinfo {author} {\bibfnamefont {L.}~\bibnamefont
  {Tarruell}}, \bibinfo {author} {\bibfnamefont {D.}~\bibnamefont {Greif}},
  \bibinfo {author} {\bibfnamefont {T.}~\bibnamefont {Uehlinger}}, \bibinfo
  {author} {\bibfnamefont {G.}~\bibnamefont {Jotzu}}, \ and\ \bibinfo {author}
  {\bibfnamefont {T.}~\bibnamefont {Esslinger}},\ }\href {\doibase
  10.1038/nature10871} {\bibfield  {journal} {\bibinfo  {journal} {Nature}\
  }\textbf {\bibinfo {volume} {483}},\ \bibinfo {pages} {302} (\bibinfo {year}
  {2012})}\BibitemShut {NoStop}%
\bibitem [{\citenamefont {Gomes}\ \emph {et~al.}(2012)\citenamefont {Gomes},
  \citenamefont {Mar}, \citenamefont {Ko}, \citenamefont {Guinea},\ and\
  \citenamefont {Manoharan}}]{Manoharan}%
  \BibitemOpen
  \bibfield  {author} {\bibinfo {author} {\bibfnamefont {K.~K.}\ \bibnamefont
  {Gomes}}, \bibinfo {author} {\bibfnamefont {W.}~\bibnamefont {Mar}}, \bibinfo
  {author} {\bibfnamefont {W.}~\bibnamefont {Ko}}, \bibinfo {author}
  {\bibfnamefont {F.}~\bibnamefont {Guinea}}, \ and\ \bibinfo {author}
  {\bibfnamefont {H.~C.}\ \bibnamefont {Manoharan}},\ }\href {\doibase
  10.1038/nature10941} {\bibfield  {journal} {\bibinfo  {journal} {Nature}\
  }\textbf {\bibinfo {volume} {483}},\ \bibinfo {pages} {306} (\bibinfo {year}
  {2012})}\BibitemShut {NoStop}%
\bibitem [{\citenamefont {Polini}\ \emph {et~al.}(2013)\citenamefont {Polini},
  \citenamefont {Guinea}, \citenamefont {Lewenstein}, \citenamefont
  {Manoharan},\ and\ \citenamefont {Pellegrini}}]{Polini}%
  \BibitemOpen
  \bibfield  {author} {\bibinfo {author} {\bibfnamefont {M.}~\bibnamefont
  {Polini}}, \bibinfo {author} {\bibfnamefont {F.}~\bibnamefont {Guinea}},
  \bibinfo {author} {\bibfnamefont {M.}~\bibnamefont {Lewenstein}}, \bibinfo
  {author} {\bibfnamefont {H.~C.}\ \bibnamefont {Manoharan}}, \ and\ \bibinfo
  {author} {\bibfnamefont {V.}~\bibnamefont {Pellegrini}},\ }\href {\doibase
  10.1038/nnano.2013.161} {\bibfield  {journal} {\bibinfo  {journal} {Nat.
  Nano.}\ }\textbf {\bibinfo {volume} {8}},\ \bibinfo {pages} {625} (\bibinfo
  {year} {2013})}\BibitemShut {NoStop}%
\bibitem [{\citenamefont {Montambaux}\ \emph {et~al.}(2009)\citenamefont
  {Montambaux}, \citenamefont {Pi\'echon}, \citenamefont {Fuchs},\ and\
  \citenamefont {Goerbig}}]{Montambaux09a}%
  \BibitemOpen
  \bibfield  {author} {\bibinfo {author} {\bibfnamefont {G.}~\bibnamefont
  {Montambaux}}, \bibinfo {author} {\bibfnamefont {F.}~\bibnamefont
  {Pi\'echon}}, \bibinfo {author} {\bibfnamefont {J.-N.}\ \bibnamefont
  {Fuchs}}, \ and\ \bibinfo {author} {\bibfnamefont {M.~O.}\ \bibnamefont
  {Goerbig}},\ }\href {\doibase 10.1103/PhysRevB.80.153412} {\bibfield
  {journal} {\bibinfo  {journal} {Phys. Rev. B}\ }\textbf {\bibinfo {volume}
  {80}},\ \bibinfo {pages} {153412} (\bibinfo {year} {2009})}\BibitemShut
  {NoStop}%
\bibitem [{\citenamefont {Bellec}\ \emph {et~al.}(2013)\citenamefont {Bellec},
  \citenamefont {Kuhl}, \citenamefont {Montambaux},\ and\ \citenamefont
  {Mortessagne}}]{Montambaux13}%
  \BibitemOpen
  \bibfield  {author} {\bibinfo {author} {\bibfnamefont {M.}~\bibnamefont
  {Bellec}}, \bibinfo {author} {\bibfnamefont {U.}~\bibnamefont {Kuhl}},
  \bibinfo {author} {\bibfnamefont {G.}~\bibnamefont {Montambaux}}, \ and\
  \bibinfo {author} {\bibfnamefont {F.}~\bibnamefont {Mortessagne}},\ }\href
  {\doibase 10.1103/PhysRevLett.110.033902} {\bibfield  {journal} {\bibinfo
  {journal} {Phys. Rev. Lett.}\ }\textbf {\bibinfo {volume} {110}},\ \bibinfo
  {pages} {033902} (\bibinfo {year} {2013})}\BibitemShut {NoStop}%
\bibitem [{\citenamefont {Feilhauer}\ \emph {et~al.}(2015)\citenamefont
  {Feilhauer}, \citenamefont {Apel},\ and\ \citenamefont
  {Schweitzer}}]{Feilhauer}%
  \BibitemOpen
  \bibfield  {author} {\bibinfo {author} {\bibfnamefont {J.}~\bibnamefont
  {Feilhauer}}, \bibinfo {author} {\bibfnamefont {W.}~\bibnamefont {Apel}}, \
  and\ \bibinfo {author} {\bibfnamefont {L.}~\bibnamefont {Schweitzer}},\
  }\href {http://arxiv.org/abs/1508.03189v1} {\bibfield  {journal} {\bibinfo
  {journal} {arXiv:1508.03189}\ } (\bibinfo {year} {2015})}\BibitemShut
  {NoStop}%
\bibitem [{\citenamefont {Katsnelson}(2012)}]{Katsnelson}%
  \BibitemOpen
  \bibfield  {author} {\bibinfo {author} {\bibfnamefont {M.~I.}\ \bibnamefont
  {Katsnelson}},\ }\href@noop {} {\emph {\bibinfo {title} {Graphene: Carbon in
  Two Dimensions}}}\ (\bibinfo  {publisher} {Cambridge University Press},\
  \bibinfo {address} {Cambridge, UK},\ \bibinfo {year} {2012})\BibitemShut
  {NoStop}%
\bibitem [{\citenamefont {Oliva-Leyva}\ and\ \citenamefont
  {Naumis}(2015{\natexlab{b}})}]{Our15b}%
  \BibitemOpen
  \bibfield  {author} {\bibinfo {author} {\bibfnamefont {M.}~\bibnamefont
  {Oliva-Leyva}}\ and\ \bibinfo {author} {\bibfnamefont {G.~G.}\ \bibnamefont
  {Naumis}},\ }\href {\doibase
  http://dx.doi.org/10.1016/j.physleta.2015.05.039} {\bibfield  {journal}
  {\bibinfo  {journal} {Physics Letters A}\ }\textbf {\bibinfo {volume}
  {379}},\ \bibinfo {pages} {2645 } (\bibinfo {year}
  {2015}{\natexlab{b}})}\BibitemShut {NoStop}%
\bibitem [{\citenamefont {Volovik}\ and\ \citenamefont
  {Zubkov}(2014)}]{Volovik14a}%
  \BibitemOpen
  \bibfield  {author} {\bibinfo {author} {\bibfnamefont {G.}~\bibnamefont
  {Volovik}}\ and\ \bibinfo {author} {\bibfnamefont {M.}~\bibnamefont
  {Zubkov}},\ }\href {\doibase http://dx.doi.org/10.1016/j.aop.2013.11.003}
  {\bibfield  {journal} {\bibinfo  {journal} {Annals of Physics}\ }\textbf
  {\bibinfo {volume} {340}},\ \bibinfo {pages} {352 } (\bibinfo {year}
  {2014})}\BibitemShut {NoStop}%
\bibitem [{\citenamefont {Volovik}\ and\ \citenamefont
  {Zubkov}(2015)}]{Volovik15}%
  \BibitemOpen
  \bibfield  {author} {\bibinfo {author} {\bibfnamefont {G.}~\bibnamefont
  {Volovik}}\ and\ \bibinfo {author} {\bibfnamefont {M.}~\bibnamefont
  {Zubkov}},\ }\href {\doibase http://dx.doi.org/10.1016/j.aop.2015.03.005}
  {\bibfield  {journal} {\bibinfo  {journal} {Annals of Physics}\ }\textbf
  {\bibinfo {volume} {356}},\ \bibinfo {pages} {255 } (\bibinfo {year}
  {2015})}\BibitemShut {NoStop}%
\bibitem [{\citenamefont {Hasegawa}\ \emph {et~al.}(2006)\citenamefont
  {Hasegawa}, \citenamefont {Konno}, \citenamefont {Nakano},\ and\
  \citenamefont {Kohmoto}}]{Hasegawa}%
  \BibitemOpen
  \bibfield  {author} {\bibinfo {author} {\bibfnamefont {Y.}~\bibnamefont
  {Hasegawa}}, \bibinfo {author} {\bibfnamefont {R.}~\bibnamefont {Konno}},
  \bibinfo {author} {\bibfnamefont {H.}~\bibnamefont {Nakano}}, \ and\ \bibinfo
  {author} {\bibfnamefont {M.}~\bibnamefont {Kohmoto}},\ }\href {\doibase
  10.1103/PhysRevB.74.033413} {\bibfield  {journal} {\bibinfo  {journal} {Phys.
  Rev. B}\ }\textbf {\bibinfo {volume} {74}},\ \bibinfo {pages} {033413}
  (\bibinfo {year} {2006})}\BibitemShut {NoStop}%
\bibitem [{\citenamefont {Vozmediano}\ \emph {et~al.}(2010)\citenamefont
  {Vozmediano}, \citenamefont {Katsnelson},\ and\ \citenamefont
  {Guinea}}]{Vozmediano}%
  \BibitemOpen
  \bibfield  {author} {\bibinfo {author} {\bibfnamefont {M.~A.~H.}\
  \bibnamefont {Vozmediano}}, \bibinfo {author} {\bibfnamefont {M.~I.}\
  \bibnamefont {Katsnelson}}, \ and\ \bibinfo {author} {\bibfnamefont
  {F.}~\bibnamefont {Guinea}},\ }\href {\doibase
  http://dx.doi.org/10.1016/j.physrep.2010.07.003} {\bibfield  {journal}
  {\bibinfo  {journal} {Physics Reports}\ }\textbf {\bibinfo {volume} {496}},\
  \bibinfo {pages} {109 } (\bibinfo {year} {2010})}\BibitemShut {NoStop}%
\bibitem [{\citenamefont {Ziegler}(2006)}]{Ziegler06}%
  \BibitemOpen
  \bibfield  {author} {\bibinfo {author} {\bibfnamefont {K.}~\bibnamefont
  {Ziegler}},\ }\href {\doibase 10.1103/PhysRevLett.97.266802} {\bibfield
  {journal} {\bibinfo  {journal} {Phys. Rev. Lett.}\ }\textbf {\bibinfo
  {volume} {97}},\ \bibinfo {pages} {266802} (\bibinfo {year}
  {2006})}\BibitemShut {NoStop}%
\bibitem [{\citenamefont {Ziegler}(2007)}]{Ziegler07}%
  \BibitemOpen
  \bibfield  {author} {\bibinfo {author} {\bibfnamefont {K.}~\bibnamefont
  {Ziegler}},\ }\href {\doibase 10.1103/PhysRevB.75.233407} {\bibfield
  {journal} {\bibinfo  {journal} {Phys. Rev. B}\ }\textbf {\bibinfo {volume}
  {75}},\ \bibinfo {pages} {233407} (\bibinfo {year} {2007})}\BibitemShut
  {NoStop}%
\bibitem [{\citenamefont {Gusynin}\ \emph {et~al.}(2007)\citenamefont
  {Gusynin}, \citenamefont {Sharapov},\ and\ \citenamefont
  {Carbotte}}]{Gusynin07}%
  \BibitemOpen
  \bibfield  {author} {\bibinfo {author} {\bibfnamefont {V.~P.}\ \bibnamefont
  {Gusynin}}, \bibinfo {author} {\bibfnamefont {S.~G.}\ \bibnamefont
  {Sharapov}}, \ and\ \bibinfo {author} {\bibfnamefont {J.~P.}\ \bibnamefont
  {Carbotte}},\ }\href {\doibase 10.1142/S0217979207038022} {\bibfield
  {journal} {\bibinfo  {journal} {International Journal of Modern Physics B}\
  }\textbf {\bibinfo {volume} {21}},\ \bibinfo {pages} {4611} (\bibinfo {year}
  {2007})}\BibitemShut {NoStop}%
\bibitem [{\citenamefont {Stauber}\ \emph {et~al.}(2008)\citenamefont
  {Stauber}, \citenamefont {Peres},\ and\ \citenamefont {Geim}}]{Stauber08}%
  \BibitemOpen
  \bibfield  {author} {\bibinfo {author} {\bibfnamefont {T.}~\bibnamefont
  {Stauber}}, \bibinfo {author} {\bibfnamefont {N.~M.~R.}\ \bibnamefont
  {Peres}}, \ and\ \bibinfo {author} {\bibfnamefont {A.~K.}\ \bibnamefont
  {Geim}},\ }\href {\doibase 10.1103/PhysRevB.78.085432} {\bibfield  {journal}
  {\bibinfo  {journal} {Phys. Rev. B}\ }\textbf {\bibinfo {volume} {78}},\
  \bibinfo {pages} {085432} (\bibinfo {year} {2008})}\BibitemShut {NoStop}%
\bibitem [{\citenamefont {Nair}\ \emph {et~al.}(2008)\citenamefont {Nair},
  \citenamefont {Blake}, \citenamefont {Grigorenko}, \citenamefont {Novoselov},
  \citenamefont {Booth}, \citenamefont {Stauber}, \citenamefont {Peres},\ and\
  \citenamefont {Geim}}]{Nair08}%
  \BibitemOpen
  \bibfield  {author} {\bibinfo {author} {\bibfnamefont {R.~R.}\ \bibnamefont
  {Nair}}, \bibinfo {author} {\bibfnamefont {P.}~\bibnamefont {Blake}},
  \bibinfo {author} {\bibfnamefont {A.~N.}\ \bibnamefont {Grigorenko}},
  \bibinfo {author} {\bibfnamefont {K.~S.}\ \bibnamefont {Novoselov}}, \bibinfo
  {author} {\bibfnamefont {T.~J.}\ \bibnamefont {Booth}}, \bibinfo {author}
  {\bibfnamefont {T.}~\bibnamefont {Stauber}}, \bibinfo {author} {\bibfnamefont
  {N.~M.~R.}\ \bibnamefont {Peres}}, \ and\ \bibinfo {author} {\bibfnamefont
  {A.~K.}\ \bibnamefont {Geim}},\ }\href {\doibase 10.1126/science.1156965}
  {\bibfield  {journal} {\bibinfo  {journal} {Science}\ }\textbf {\bibinfo
  {volume} {320}},\ \bibinfo {pages} {1308} (\bibinfo {year}
  {2008})}\BibitemShut {NoStop}%
\bibitem [{\citenamefont {Lee}\ \emph {et~al.}(2015)\citenamefont {Lee},
  \citenamefont {Kim}, \citenamefont {Na}, \citenamefont {Chang}, \citenamefont
  {Kim}, \citenamefont {Yu}, \citenamefont {Lee},\ and\ \citenamefont
  {Kim}}]{Seung15}%
  \BibitemOpen
  \bibfield  {author} {\bibinfo {author} {\bibfnamefont {S.-M.}\ \bibnamefont
  {Lee}}, \bibinfo {author} {\bibfnamefont {S.-M.}\ \bibnamefont {Kim}},
  \bibinfo {author} {\bibfnamefont {M.}~\bibnamefont {Na}}, \bibinfo {author}
  {\bibfnamefont {H.}~\bibnamefont {Chang}}, \bibinfo {author} {\bibfnamefont
  {K.-S.}\ \bibnamefont {Kim}}, \bibinfo {author} {\bibfnamefont
  {H.}~\bibnamefont {Yu}}, \bibinfo {author} {\bibfnamefont {H.-J.}\
  \bibnamefont {Lee}}, \ and\ \bibinfo {author} {\bibfnamefont {J.-H.}\
  \bibnamefont {Kim}},\ }\href {\doibase 10.1007/s12274-015-0719-9} {\bibfield
  {journal} {\bibinfo  {journal} {Nano Research}\ }\textbf {\bibinfo {volume}
  {8}},\ \bibinfo {pages} {2082} (\bibinfo {year} {2015})}\BibitemShut
  {NoStop}%
\end{thebibliography}%

\end{document}